\documentclass[12pt]{article}
\usepackage[dvips]{graphicx}
\usepackage{subfigure}
\usepackage{epsfig,cite}
\usepackage {amsmath}
\usepackage{amssymb}
\usepackage{slashed}
\usepackage{rotating}
\usepackage{bm}
\def\etaw{\eta_{\rm wmap}}

\def\iso#1#2{\mbox{${}^{#2}{\rm #1}$}}

\def\be#1{\iso{Be}{#1}}
\def\bor#1{\iso{B}{#1}}
\def\li#1{\iso{Li}{#1}}
\def\car#1{\iso{C}{#1}}

\def\be#1{\iso{Be}{#1}}
\def\he#1{\iso{He}{#1}}

\def\sbhe#1{\mbox{\scriptsize\he{#1}}}

\def\eres{E_{\rm res}}
\def\g#1{\Gamma_{\rm #1}}
\def \geff{\Gamma_{\rm eff}}
\def \gtot{\Gamma_{\rm tot}}

\def \sigv{\avg{\sigma v}}

\def \3heag{\avg{\sigma v}_{\rm\mbox{\scriptsize \he3}\rm\alpha}}
\def \benp{\avg{\sigma v}_{\rm\mbox{\scriptsize \be{7}}n}}

\def \beXA{\avg{\sigma v}_{\rm\mbox{\scriptsize\be{7}} X}}

\def\avg#1{\langle #1 \rangle}

\def \rk{\rm keV}

\def\beq{\begin{equation}}
\def\eeq{\end{equation}}
\def\beqa{\begin{eqnarray}}
\def\eeqa{\end{eqnarray}}
\def\beqas{\begin{eqnarray*}}
\def\eeqas{\end{eqnarray*}}
\def \l({\left(}
\def \r){\right)}

\def\pfrac#1#2{\left( \frac{#1}{#2} \right)} 

\def\ga{\mathrel{\raise.3ex\hbox{$>$\kern-.75em\lower1ex\hbox{$\sim$}}}}
\def\la{\mathrel{\raise.3ex\hbox{$<$\kern-.75em\lower1ex\hbox{$\sim$}}}}

\begin{document}

\begin{titlepage}
\pagestyle{empty}
\rightline{UMN--TH--2924/10}
\rightline{FTPI--MINN--10/31}
\rightline{Final version:  March 2011}
\begin{center}
{\large {\bf Resonant Destruction as a Possible Solution \\
to the Cosmological Lithium Problem}}
\end{center}
\begin{center}
{\bf Nachiketa Chakraborty$^{1}$, Brian D. Fields$^{1}$} and {\bf Keith A. Olive}$^{2}$
\vskip 0.2in
{\small {\it
$^1$Departments of Astronomy and of Physics, \\ University of Illinois, Urbana, IL 61801, USA\\
$^2$William I. Fine Theoretical Physics Institute, \\
University of Minnesota, Minneapolis, MN 55455, USA}}
\vskip 0.2in
{\bf Abstract}
\end{center}
{
We explore a nuclear physics resolution to the discrepancy between the predicted 
standard big-bang nucleosynthesis (BBN) abundance of \li7 and its observational determination in metal-poor stars. 
The theoretical $\li{7}$ abundance is 3 - 4 times greater than the observational values, assuming the baryon-to-photon ratio, $\etaw$, determined by WMAP. The \li7 problem could be resolved within the standard BBN picture if additional 
destruction of $A=7$ isotopes occurs due to new nuclear reaction channels or upward corrections to
existing channels. This could be achieved 
via missed resonant nuclear reactions, which is the possibility we consider here. We find some potential candidate resonances which can solve the lithium problem 
and specify their required resonant energies and widths. 
For example, a 1$^-$ or 2$^-$ excited state of $^{10}$C sitting at approximately 15.0 MeV
above its ground state with an effective width of order 10 keV could
resolve the \li7 problem;
the existence of this excited state needs experimental verification. 
Other examples using known states include
$\be7 + t \rightarrow \bor{10}(18.80 \ \rm MeV)$, and 
$\be7+d \rightarrow \bor9(16.71\ \rm MeV)$.
For all of these states, a large channel radius ($a > 10$ fm) is needed to give sufficiently
large widths.  
Experimental determination of these reaction strengths is needed to rule out or confirm
these nuclear physics solutions to the lithium problem.}



\end{titlepage}

\section{\label{sec:level1} Introduction}

Primordial nucleosynthesis continues to stand as our earliest
probe of the universe based on Standard Model physics.
Accurate estimates of the primordial abundances of 
the light elements D, \he4 and \li7 within standard Big Bang Nucleosynthesis (BBN)
\cite{cfo1,coc,bbn2,cyburt,Cyburt:2008kw} are crucial for making comparisons 
with observational determinations and ultimately testing the theory.
Primordial abundances are also a probe of the early universe physics \cite{cfos}. 
Currently,  the theoretical estimates of D and \he4 
match the observational values 
within theoretical and observational uncertainties \cite{bbn2,Cyburt:2008kw} 
at the baryon-to-photon ratio determined by the 7-year WMAP data, 
$\etaw = 6.19 \pm 0.15 \times 10^{-10}$ 
\cite{wmap7}. In contrast, the theoretical primordial 
abundance of  $\li{7}$  does not match the observations. 

At $\etaw$, the predicted BBN abundance of \li7 is\footnote{Note that the \li7 abundance reported here differ slightly from that
given in \cite{Cyburt:2008kw}, primarily due to the small shift in $\eta$ as reported in \cite{wmap7}.} \cite{Cyburt:2008kw}
\beq
\left( \frac{\li7}{\rm H} \right)_{\rm BBN} 
 = \left( 5.12^{+0.71}_{-0.62} \right) \times 10^{-10} .
\eeq
The observed \li7 abundance is derived from observations of low-metallicity halo 
dwarf stars which show a plateau\cite{Spite:1982dd}
in (elemental) lithium versus metallicity,
with a small scatter consistent with observational uncertainties.
An analysis \cite{Ryan:2000zz}
of field halo stars gives a plateau abundance of
\beq
\label{eq:li_halo}
\left( \frac{\rm Li}{{\rm H}} \right)_{\rm halo \star}
 = (1.23^{+0.34}_{-0.16}) \times 10^{-10} .
\eeq
However,
the lithium abundance in several globular clusters tends to be 
somewhat higher~\cite{liglob,new}, and
a recent result found in \cite{new} gave
\li7/H = $(2.34 \pm 0.05) \times 10^{-10}$.
Thus the theoretically estimated abundance of the isobar 
with mass 7 ($\be{7}$+$\li{7}$) is more than the observationally determined
value by a factor of 2.2 - 4.2 \cite{Cyburt:2008kw}, 
at $\etaw$. Relative to the theoretical and observational uncertainties,
this represents a deviation of 4.5-5.5 $\sigma$.

This significant discrepancy constitutes the ``lithium problem''
which could point to 
limitations in either the 
observations, our theoretical understanding of
nucleosynthesis, or the post-BBN processing of lithium. 

On the theoretical front, strategies which have
emerged to 
approach the lithium problem broadly either
address astrophysics or microphysics.
On the astrophysical side, 
one might attempt to improve our understanding of 
lithium depletion mechanisms operative in stellar models \cite{dep}.
This remains an important goal but is not our focus here.

The microphysical solutions to the lithium problem
all in some way change the nuclear reactions
for lithium production in order to reduce the primordial (or pre-Galactic)
lithium abundance to observed levels.
Some of these work within the Standard Model, focussing on nuclear
physics, in particular the nuclear reactions involved
in lithium production.  One approach is to attempt to
utilize the experimental uncertainties in 
the rates \cite{coc,coc3,cfo4,boyd}. 
A second, related approach is the inclusion of new effects in the 
nuclear reaction database such as poorly understood resonance effects \cite{Cyburt:2009cf}.
Finally, it may happen that effects beyond the Standard Model
are responsible for the observed lithium abundance.  For example, 
the primordial lithium abundance can be reduced by
cosmological variation of the fine structure constant associated with 
a variation in the deuterium binding energy  \cite{vary}, or 
by the post-BBN destruction of lithium through the late decays of a massive
particle in the early universe \cite{jed04}.

In this paper, we remain within the Standard Model, examining
the possible role of resonant reactions which may have been up to now neglected. 
The requisite reduction in the \li7 abundance 
 can be achieved by either an enhancement  in the rate of 
destruction of \li{7} or its mirror nucleus \be{7}. This approach is more 
promising than the alternative of 
reducing the production of \be{7} and \li{7} 
where the reactions are better understood experimentally and theoretically
\cite{Cyburt:2004cq,Cyburt:2008up,Ando:2005cz},
whereas the experimental and especially the theoretical situation
for $A = 8-11$ has made large strides but still allows
for surprises at the levels of interest to us \cite{qmc}. 

The use of resonant channels is an approach that has 
paid off in the past in the context of stellar nucleosynthesis. 
Fred Hoyle  famously predicted a 
resonant energy level at 7.68 MeV  in the $\car{12}$ compound 
nucleus which enhances the $\be{8} + \alpha \rightarrow \car{12}$ 
reaction cross-section and allows the triple alpha reaction 
to proceed at relatively low densities \cite{Hoyle:1954zz}. 
Recently, it was shown that there are promising 
resonant destruction mechanisms which can achieve 
the desired reduction of the total $A=7$ isotopic abundance \cite{Cyburt:2009cf}. 
This paper points to a resonant energy level at 
$(E,J^\pi) = (16.71 \ {\rm MeV},5/2^{+})$ in the $\bor{9}$ compound 
nucleus which can increase the rate of the \be{7}(d,p)$\alpha \alpha$ and/or
$\be{7}(d,\gamma)\bor{9}$ and thereby reduce 
the $\be{7}$ abundance. 
Here, we take a more general approach 
and systematically search for all possible
compound nuclei \cite{tunl} and potential resonant channels 
 which may result in the destruction of \be7 and/or \li7.

Because of the large discrepancy between the observed and BBN abundance of \li7,
any nuclear solution to the lithium problem will require 
a significant modification to the existing rates.
As we discuss
in the semi-analytical estimate in section~\ref{semi-analytical}, 
any new rate or modification to an existing one, must be  
2 - 3 times greater than the current dominant destruction channels namely,
$\li{7}(p,\alpha)\alpha$ for \li{7} and 
$\be{7}(n,p)\li{7}$ for \be{7}. 
As discussed in \cite{boyd} and as we show semi-analytically
in \S~\ref{semi-analytical}, this is difficult to achieve
with non-resonant reactions. Hence, we will concentrate on 
possible resonant reactions as potential
solutions to the lithium problem. 
As we will show, there are 
interesting candidate resonant channels which 
may resolve the \li7 problem.   For example, there is a 
possibility of 
 destroying \be7 through a 1$^-$ or 2$^-$ $^{10}$C excited state at approximately 15.0 MeV.
 The energy range between 6.5 and 16.5 MeV is currently very poorly mapped out
 and a state near the entrance energy for \be7 + \he3 could provide a  solution
if the effective width is of order 10 keV. 
 We will also see that these
reactions all require fortuitously favorable nuclear parameters,
in the form of large channel radii, as also found by
Cyburt and Pospelov \cite{Cyburt:2009cf} in the case of 
$\be7+d$.
Even so, in the face of the more radical alternative of new fundamental particle
physics, these more conventional solutions to the lithium problem
beckon for experimental testing.

The paper is organized as follows: First, we
lay down the required range of properties of 
any resonance to
solve the lithium problem by means of a semi-analytic estimate 
inspired by \cite{Mukhanov:2003xs, Esmailzadeh:1990hf} in \S~\ref{semi-analytical}. 
Then, in \S~\ref{sect:search}, we list experimentally identified resonances from the 
databases: TUNL \cite{tunl} and NNDC \cite{nndc}, 
involving either the destruction of
$\be{7}$ or $\li{7}$. Finally, the 
solution space of resonant properties, wherein the lithium problem
is partially or completely solved, is mapped for the most promising
initial states involving either \li{7} or \be{7}, 
by including these rates in a numerical estimation of
the \li{7} abundance. This exercise will delineate the 
effectiveness of experimentally studied or identified 
resonances as well as requirements of possible missed resonant 
energy levels in compound nuclei formed by these initial states. 
 This is described in \S~\ref{numerical}. We note that in our 
analysis, the narrow resonance approximation is assumed which may
not hold true in certain regions of this solution space.
Our key results are pared down to a few resonant reactions described
in \S~\ref{reducedlist}.  A summary and conclusions are given in \S~\ref{disc}.

\section{\label{semi-analytical} Semi-analytical estimate of important reaction rates}

Before we embark on a systemic survey of possible resonant enhancements
of the destruction of $A = 7$ isotopes, it will be useful
to estimate the degree to which the destruction rates must change 
in order to have an impact on the final \li7 abundance.
The net rate of production of a nuclide $i$ is given by the 
difference between the production from nuclides $k$ and $l$ 
and the destruction rates via nuclide $j$, i.e. for the reaction ${ i + j \rightarrow k + l}$. 
This is 
expressed quantitatively by the rate equation \cite{Wagoner:1966pv}
for abundance change
\beqa
\label{eq:numraterxn}
\frac{d n_{i}}{dt} 
  =  -3H n_i + \sum_{jkl} n_{k} n_{l} \avg{\sigma v}_{kl} 
    - n_{i} n_{j} \avg{\sigma v}_{ij}  ,
\eeqa
where $ n_{i}$ is the number density of nuclide $i$,  $H$ is the Hubble parameter, 
${ \sum_{ij} n_{i} n_{j} \avg{\sigma v}_{ij}}$ are 
the sum of contributions from all the forward reactions
destroying nuclide $i$ and 
${ \sum_{kl} n_{k} n_{l} \avg{\sigma v}_{kl} }$ are the 
reverse reactions producing it. ${\rm \avg{\sigma v} }$ is the 
thermally averaged cross-section of the reaction. 
The dilution of the density 
of these nuclides due 
to the expansion of the universe can be removed by re-expressing 
eq.~(\ref{eq:numraterxn}) in terms of 
number densities relative to the baryon density
$Y_i \equiv n_i/n_{\rm b}$, as,
\beqa
\label{eq:raterxn}
\frac{d Y_{i}}{dt}  
  &=&  n_{\rm b} \sum_{jkl}  Y_{k} Y_{l} \avg{\sigma v}_{kl} 
           - Y_{i} Y_{j} \avg{\sigma v}_{ij}   \ \ .
\eeqa
Using this general form, the net rate of \be{7} production can 
be approximated in terms of the thermally averaged cross-sections 
of its most important production and destruction channels as
\beqa
\label{eq:berateeqn}
\frac{d Y_{ \mbox{\scriptsize \be7}}}{dt} 
  =  n_{\rm b}\ ( \3heag Y_{\mbox{\scriptsize\he3}}  Y_{\alpha} 
        - \benp Y_{\mbox{\scriptsize \be7}}  Y_{n} ) \ \ .
\eeqa
Here, the reverse reaction rates of these production and destruction channels 
are neglected, as they are much smaller than the forward rates 
at the lithium synthesis temperature.
 A similar equation 
can be written down for \li{7}.
When quasi-static equilibrium is reached, the 
destruction and production rates are equal. In this case, 
approximate values for new rates, 
which can effectively destroy either isobar, can be obtained analytically.

At temperatures $\rm T \approx 0.04$ MeV, both \li{7} and 
\be{7} are in equilibrium \cite{Mukhanov:2003xs} which gives,
\beq\label{equilibrium}
\3heag Y_{\rm \mbox{\scriptsize\he{3}}}   Y_{\rm \alpha}  
  = \benp Y_{\rm \mbox{\scriptsize\be{7}}}  Y_{n}  \ \ .
\eeq
Consider a new, inelastic \be7 destruction channel $\be{7} + X \rightarrow Y + Z$,
involving projectile $X$. 
This reaction will add to the right hand 
side of eq.~(\ref{equilibrium}) and
shift the equilibrium abundance of \be{7} to a new value as follows,
\beq
\label{eq:newLi}
Y_{\mbox{\scriptsize\be{7}}}^{\rm new} 
  \approx \frac{ {\rm \3heag} Y_{\rm \alpha} Y_{\rm\sbhe3}  }{{\rm \benp} Y_{\rm n} + \beXA Y_{X} } \\ 
   \approx \frac{1}{1 \ +\ \frac{ \beXA Y_{X}}{ \benp Y_{n}}} 
  \ Y_{\mbox{\scriptsize\be{7}}}^{\rm old} \  \ .
\eeq
If the new reaction is to be important in solving the lithium problem,
it must reduce the \be{7} abundance by a factor of 
$Y_{\mbox{\scriptsize\be{7}}}^{\rm new}/Y_{\mbox{\scriptsize\be{7}}}^{\rm old} \sim 3 - 4$ .
This in turn demands via eq.~(\ref{eq:newLi}) that 
$\beXA Y_{X}/\benp Y_{n} \sim 2 - 3$, i.e., the rate for the new reaction
exceeds that of the usual $n-p$ interconversion reaction rate. A similar estimate 
can be made for  \li{7}. 

This reasoning would exclude 
non-resonant rates as they would be required to have unphysically
large astrophysical $S$-factors in the range of
order $10^{5} - 10^{9}$ \rm keV - barn depending on the channel. 
Thus we would expect that only resonant reactions can produce
the requisite high rates. 
Possible resonant reactions are listed in the next section,
whose key properties 
of resonance strength, $\geff$ and
energy, $\rm E_{\rm res}$, lie in appropriate ranges capable of
 achieving the required destruction of 
mass 7. 

Finally, we turn to \li7 destruction reactions,
$\li7+X \rightarrow Y + Z$.
Recall that at the WMAP value of $\eta$, mass 7 is made predominantly
as \be7, with direct \li7 production about an order of
magnitude smaller.  This suggests that enhancing 
direct \li7 destruction will only modestly affect the 
final mass-7 abundance; we will see that this expectation is largely
correct.\footnote{
A subtle point is that normally, the mass-7 abundance
is most sensitive to 
rate $\be7(n,p)\li7$ \cite{skm}.
Of course, this reaction leaves the mass-7 abundance unchanged,
but the lower Coulomb barrier for \li7
leaves it vulnerable to the
$\li7(p,\alpha)\he4$ reaction,
which is extremely effective in removing \li7.
Thus, for a new, resonant \li7 destruction reaction to be
important, it must successfully compete with the very large 
$\li7(p,\alpha)\he4$ rate, and even then the mass-7 destruction
``bottleneck'' remains  the $\be7(n,p)\li7$ rate
that limits \li7 appearance.
Thus we would not expect direct \li7 destruction
to be effective.  We will examine \li7 destruction
below, and confirm these expectations.}

With these pointers, 
the list in the next section 
is reduced and numerical analysis of the 
remaining promising rates is done.

\section{Systematic Search for Resonances
\label{sect:search} 
}

In this section we describe a systematic search for
nuclear resonances which could affect primordial lithium production.
We first begin with general considerations, then catalog
the candidate resonances.
We briefly review the basic physics of resonant reactions 
to establish notation and highlight the key physical ingredients.

\subsection{General Considerations
\label{sect:general} }

Energetically,
the net process  $\be7 + A \rightarrow B + D$
must have $Q+E_{\rm init}\ge 0$,
where the initial kinetic
energy $E_{\rm init} \simeq T \la 40$ keV
is small at the epoch of $A=7$ formation. 
Thus we in practice require
exothermic reactions, $Q>0$.  Moreover,
inelastic reactions with large $Q$
will yield final state particles with large
kinetic energies.  Such final states thus have larger
phase space than those with small $Q$ and in that
sense should be favored.

Consider now a process $\be7 + X \rightarrow C^* \rightarrow Y + Z$
which destroys \be7 via a resonant compound state;
a similar expression can be written for \li7 destruction.
In the entrance channel $\be7 + X \rightarrow C^*$ 
the energy released in producing the compound state is
$Q_C = \Delta(\be7) + \Delta(X) - \Delta(C^{\rm g.s.})$,
where $\Delta(A) = m-Am_{\rm u}$ is the mass defect.
If an excited state $C^*$ in the compound nucleus
lies at energy $E_{\rm ex}$, then the difference
\beq
\eres \equiv E_{\rm ex} - Q_C 
\label{ex}
\eeq
determines the effectiveness of the resonance.
We can expect resonant 
production of $C^*$ if 
$\eres \la T$.
In an ordinary (``superthreshold'') resonance
we then have $\eres > 0$, while
a subthreshold resonance has
$\eres < 0$.

Once formed, the excited $C^*$ level can decay via 
some set of channels.
The cross section for $\be7 + X \rightarrow C^* \rightarrow Y + Z$
is given by the Breit-Wigner expression
\beq
\label{eq:bw}
\sigma(E) = \frac{\pi \omega}{2\mu E} 
   \frac{\Gamma_{\rm init}\Gamma_{\rm fin}}
        { (E-\eres)^2 + (\Gamma_{\rm tot}/2)^2 }
\eeq
where $E$ is the center-of-mass kinetic energy in the initial state, $\mu$ is the 
reduced mass and 
\beq
\label{eq:omega}
\omega = \frac{2J_{C^*}+1}{(2J_{X}+1)(2J_{7}+1)}
\eeq
is a statistical factor accounting for angular momentum.
The width of the initial state (entrance channel)
is $\Gamma_{\rm init}$, and the width of the  
final state (exit channel) is $\Gamma_{\rm fin}$.

One decay channel which must always be available is the
entrance channel itself.  Obviously such an elastic reaction
is useless from our point of view.
Rather, we are interested in inelastic reactions in which
the initial \be7 (or \li7) is transformed to something else.
In some cases, an inelastic strong decay is possible
where the final state particles $Y+Z$ are both nuclei.
Note that it is possible to produce a final-state nucleus in an excited
state, e.g., $C^* \rightarrow Y^* + Z$, in which case
the energy release $Q^\prime_C$ is offset by the $Y^*$ excitation energy.
This possibility increases the chances of finding energetically
allowable final states.  Indeed, such a possibility 
has been suggested in connection with the 
$\be7 + d \rightarrow \bor{9}^* \rightarrow \be8^* + p$  process
\cite{Cyburt:2009cf}.

Regardless of the availability of a strong inelastic channel,
an electromagnetic transition $C^* \rightarrow C^{(*)} + \gamma$
to a lower level is always possible. However, these often 
have small widths and thus 
a small branching ratio $\Gamma_{\rm fin}/\Gamma_{\rm tot}$.
Thus for electromagnetic decays to be important, 
a strong inelastic decay must not be available, and the
rest of the reaction cross section needs to be large to
compensate the small branching; as seen in eq.~(\ref{eq:bw}),
this implies that $\Gamma_{\rm init}$ be large.

Note that in all charged-particle
reactions, the Coulomb barrier is crucially important and is implicitly
encoded via the usual exponential Gamow factor in the reaction widths
of both initial and final charged-particle states. 
However, if the reaction has a high $Q$, 
the final state kinetic energy will be large and thus there will
not be significant final-state Coulomb supression;
this again favors final states with large $Q$.
In addition, if the entrance or exit channel 
has orbital angular momentum $L > 0$, there is additional
exponential suppression, so that $L > 0$ states are disfavored
for our purposes.

With these requirements in mind, we will systematically 
search for resonant reactions which could ameliorate or solve
the lithium problem.  We begin by identifying possible
processes which are
\begin{enumerate}

\item
{\em new} resonances
not yet included in the BBN code;

\item
2-body to 2-body processes, since 3-body rates are generally very
small in BBN due to phase space suppression as well as the 
relatively low particle densities and short timescales; 

\item
experimentally allowed -- in practice this means we seek
unidentified states in poorly studied regimes; 

\item
{\em narrow} resonances having $\gtot \la T$, which
is around $\gtot < 40$ keV but we will also consider somewhat
larger widths to be conservatively generous.

\item
relatively low-lying resonances with $\eres \la {few} \times T \sim 100-300$ 
keV,
which are thermally accessible; here again we err on the side of a generous
range.

\end{enumerate}
Once we have identified all possible candidate resonances, 
we will then assess their viability as solutions to the lithium problem
based on available nuclear data.

\subsection{\label{list} List of Candidate Resonances}

As described above, 
we will explore the 
resonant destruction channels of both $\li{7}$ and $\be{7}$. 
Some of the 
potential resonances which might be able to reduce 
the mass 7 abundance to the observed value were recently considered in \cite{Cyburt:2009cf}. 
This analysis eliminates several candidate 
resonances, leaving as genuine solutions
only the resonance related to the $\be{7}(d,\gamma)\bor{9}$ 
and $\be{7}(d,p)\alpha\alpha$ reactions and associated with the
16.71 MeV level in the $\bor{9}$ compound nucleus. Here, we make
an exhaustive list of possible promising resonances that 
may be important to either $\be{7}$ or $\li{7}$ destruction 
channels. In order to do so systematically and 
account for all possible resonances that may be of 
importance, we study the energy levels in all possible 
compound nuclei that may be formed in destroying 
$\be{7}$ or $\li{7}$, making extensive use of databases
at TUNL and the NNDC \cite{tunl,nndc}. 

The available 2-body destruction channels 
${}^{7}A + X$ may be classified by
$X= n, p, d, t, \he{3}, \alpha$, and $\gamma$. 
Consequently, the compound nuclei that can be formed 
starting from mass 7 have mass numbers ranging 
from $A=8$ to $A=11$, and the ones of particular interest 
are $\li{8}, \be{8}, \bor{8}, \be{9}, \bor{9}, 
\be{10}, \bor{10}, \car{10}, \bor{11}$ and $\car{11}$. 
All relevant, resonant energy levels in these compound 
nuclei that may provide paths for reduction of 
mass-7 abundance 
are listed in Tables~\ref{table:candidatesbebor8} --  \ref{table:candidatescar10bor11car11}. 

There are 
quantum mechanical and kinematic restrictions to 
our selection of candidates. 
The candidate 
resonant reactions must obey selection rules. The 
partial widths for a channel, which may be viewed as probability currents of 
emission of the particle in that channel through the nuclear surface, 
are given as 
\beqa
\label{eq:partialwidth}
\Gamma_{L}(E) = 2\ k a\ P_{L}(E,a) \gamma^{2}(a)
\eeqa
where $a$ is the channel radius and $E$ is the projectile 
energy.
Here 
$k$ is the wavenumber of the colliding particles in 
the centre-of-mass frame and $\gamma^{2}$ is 
the reduced width, which depends  
on the overlap between the wavefunctions inside 
and outside the nuclear surface, 
beyond which the nuclear forces are unimportant. 
The reduced width, $\gamma^{2}$ is independent of energy and has a 
statistical upper limit called the Wigner limit 
given by \cite{Teichmann:1952}
\beq
\label{eq:TeichmannWignerlimit}
\gamma^{2} \leq \frac{3 \hbar^2}{2 \mu a^{2}} ,
\eeq
The pre-factor of $\frac{3}{2}$ is under the assumption that 
the nucleus is uniform and can change to within a factor 
of order unity if this assumption changes. 
The Wigner limit depends sensitively on the 
channel radius and thus varies with the 
nuclei involved. For the nuclei of our interest, 
typical values of $\gamma^{2}$ range from 
a few hundred {\rm keVs} to a few {\rm MeVs}. 

In eq.~(\ref{eq:partialwidth}),
$P_{L}(E,a)$ is the Coulomb penetration probability for angular momentum $L$ 
and is a strong and somewhat complicated function of $E$ and $a$. 
Thus, while the Wigner limit sets a theoretical limit on 
the reduced width, the upper limit on the full width, $\Gamma_{L}(E)$, 
depends on the values of $P_{L}(E,a)$ and 
is sensitive to the details of the resonant channel 
being considered. In light of this complexity, our strategy 
is as follows. We evaluate the $\Gamma_{L}(E)$ needed
to make a substantial impact on the lithium problem.
Then for the cases of highest interest, we will compare
our results with the theoretical limit set by the Coulomb 
suppressed Wigner limit for those specific cases. 

We also limit our consideration to two body initial states, with 
resonance energies $\eres \le 650$\ keV. The 
high resonance energy limit ensures that all possible 
resonances which may influence the final \li7 abundance 
are taken into account, though many of the channels with 
such high resonance energies will inevitably 
be eliminated. Excited final states have also been 
considered in making this list. Different excited states 
of final state products are marked as separate entries 
in the table, since each one has its own spin and 
therefore a different angular momentum barrier. And 
thereby the significance of each excited state 
in destroying mass 7 is varied. Also, we usually 
eliminate the reactions with a negative Q-value 
except for the $\li7(d,p)\li8$, $\be7(d,\he3)\li6$, 
$\be7(d,p)\be8^{*}$ (16.922 MeV) and $\li7(\he3,p)\be9^{*}$ (11.283 MeV)
as they are only marginally endothermic.

For a number of the reactions listed in these tables, 
1-10, the total spin of the initial state reactants is equal 
to that of the compound nucleus, which is equal to the total spin 
of the products, with $L = 0$. 
However, for many reactions, 
angular momentum is required in the initial and/or final state, which 
decreases the penetration probability and thereby 
the width for that particular channel. In fact 
for some of these reactions, parity 
conservation demands higher angular momentum 
which worsens this effect. However, we do not reject 
any channel based on the angular momentum 
suppression of its width 
in these tables. 
Later we will shortlist those resonant channels 
which are not very suppressed and 
indeed potentially effective 
in destroying mass 7.

The reactions of interest are listed in increasing 
order of the mass of the compound nuclei formed. The particular resonant 
energy levels of interest $E_{\rm ex}$ and their spins 
are listed in the table. In general, different initial states involving 
\li7 and \be7 can form these energy levels and so all these relevant initial states 
are listed. For each one, the various final product states for an inelastic reaction 
are enumerated. Again, each of the final state products can also be formed 
in an excited state. These excited states must have lower energy 
than the initial state energies for the reaction to be exothermic. 
In addition spin and parity must be conserved. Enforcing these, 
the minimum final state angular momenta $L_{\rm fin}$ are evaluated from the 
spin of the resonant energy level and are listed in the tables. 
The total widths of the energy levels are listed whenever available.
The partial widths of the different channels including the elastic one, 
out of each energy level 
are also listed.

We adopt the narrow resonance approximation 
to evaluate the effect of these resonances and 
either retain or dismiss them as potential solutions to the lithium problem.
Some of the partial widths or limits on them 
are high enough that they easily 
qualify to be broad resonances. 
This implies that the narrow resonance formula used to see 
their effect is not precise, but still gives a rough idea 
of whether the resonance is ineffective or 
not.

Our expression for thermonuclear rates in the narrow
resonance approximation is explained in detail in
Appendix \ref{sect:narrow}, and is given by
\beqa
\label{eq:analres}
\avg{\sigma v} &=& 
\omega \ \geff \left(\frac{2\pi}{\mu T}\right)^{3/2} 
  e^{-|\eres|/T} \ f\left( 2\eres/\gtot \right)
\eeqa
This rate is controlled by two parameters specific
to the compound nuclear state:  $\eres$ and $\geff$.
Here $\eres$ is given in eq.~(\ref{ex}),
and measures the offset from the entrance channel and 
the compound state.  
The resonance strength is quantified via
\beq
\geff = \frac{\g{init} \g{fin}}{\gtot}
\eeq
with $\g{init}$ and 
$\g{fin}$ being the entrance and exit widths of a 
particular reaction, and $\gtot$ the sum of the
widths of all possible channels. Of these widths, 
the smaller of $\g{init}$ and $\g{fin}$ 
along with $\gtot$ are listed in the table above. 
The resonance strength, $\geff \approx \rm \g{init}$, if
$\g{fin}$ dominates the total width, and vice versa. 
If $\g{init}$ and $\g{fin}$ are the dominant partial 
widths and they are comparable to each other, 
then the strength is even higher. 

As discussed in Appendix \ref{sect:narrow}, 
our narrow resonance rate in eq.~(\ref{eq:analres})
improves on the form of the usual expression
for narrow resonance in two ways: (a) it extends to the
subthresold domain; and (b) it introduces
the factor $f$ which accounts for
a finite $\eres/\gtot$ ratio.

It is important to make a systematic and 
comprehensive search for all possible experimentally identified 
resonances capable of removing this discrepancy. 
In addition, it is possible that 
resonances and indeed energy levels themselves were missed, 
especially at the higher energies, where uncertainties are greater. 
Therefore it is useful to map the parameter space where the 
lithium discrepancy is removed to apriori lay down 
our expectations of such missed resonances. 
This can be done by looking at interesting 
initial states involving \li{7} and \be{7}, and abundant 
projectiles $p, n, d, t, \he3, \alpha$,
and parametrizing the effect of inelastic channels on the 
mass-7 abundance. This is described in
\S~\ref{numerical}. 

\section{\label{numerical}Narrow resonance solution space}

In order to study the effect of resonances in different 
compound nuclei on the abundance of mass 7, 
our strategy is to specify the reaction rate for possible
resonances,
and then run the BBN code to find the mass-7 abundance
in the presence of these resonances.
In particular, for reactions involving light projectile $X$,
we are interested in considering the general effect of
states 
${}^{7}A + X \rightarrow C^*$, including those associated
with known energy levels in the compound nucleus,
as well as possible overlooked states.

We 
assume that the narrow resonance approximation 
holds true at least as a rough guide.  
If the reaction pathway is specified, i.e.,
all of the nuclei ${}^{7}A + X \rightarrow C^* \rightarrow Y + Z$
are identified, then 
the reduced mass $\mu$, 
reverse ratio 
and the $Q$-value
are uniquely determined.
In this case, the thermally averaged 
cross-section is given by eq.~(\ref{eq:analres}),
with two free parameters: the product $\omega \geff$ and
the resonance energy, $\eres$. 
Because the state $C^*$ is unspecified,
so is its spin $J_*$.  On the other hand, we
do know the spins of the initial state particles,
and thus $\omega$ is specified up to a factor $2J_*+1$
(eq.~\ref{eq:omega}).
For this reason, the 
$\omega \geff$ dependence reduces to
$(2J_*+1)\geff$, which we explicitly indicate in all of our plots.

In a few cases we will be interested 
in one specific final quantum state, 
 e.g., $\be7(t,\he3)\li7$; 
when the final state is specified,
the reaction can be completely determined,
including the effect of the reverse rates.
However, in most situations we are interested
in the possibility of an overlooked excited state
in the compound nucleus, and thus 
in unknown final states.  In this scenario
we thus have only a ``generic'' 
inelastic exit channel.  Consequently,
for such plots we cannot evaluate the reverse reaction rate
(which is in all interesting cases small)
and so we set the reverse ratio to zero.

The resonant rates are included in the BBN code, 
individually for compound nuclei with an interesting 
initial state. The plots below show contours of constant, reduced mass-7 
abundances. A general feature of all the plots, is 
the near linear relation between $\log \geff$ 
and $\eres$ in the region of larger, positive values 
of $\geff$ and $\eres$. This can seen 
quantitatively as follows. The thermal rate is integrated 
over time or equivalently temperature to give 
the final abundance of mass 7 or \li{7} as it exists. 
Now assuming that the thermal rate operates at an 
effective temperature, $T_{\rm Li}$, at which 
$\li{7}$ production peaks, a given value for this effective $\sigv$ 
will give a fixed abundance. This implies,
\beqa
\delta Y_7/Y_7 \sim \avg{\sigma v}_{\rm peak} \sim \geff \ e^{-\eres/ T_{\rm Li}} \sim \ \rm constant
\eeqa
This gives a feel for the linear relation in the plot.

\subsection{\label{compnuc8} A = 8 Compound Nucleus}

\begin{table}[h!]
\begin{center}
\rotatebox{0}{
\scalebox{0.78}{
\begin{tabular}{|c|c|c|c|c|c|c|c|}
\hline
\hline
Compound Nucleus, & Initial & $L_{\rm init}$ & $L_{\rm fin}$ & $E_{\rm res}$ & $\gtot$ & \rm Exit & \rm Exit Channel \\ 
$J^\pi, E_{\rm ex}$ & State & & & & & Channels & Width \\
\hline
\li{8}, $3^{+}$, 2.255\ \rm MeV & $\li{7} + n$ & 1 & 1 & 222.71 \rm{keV} & $33\pm6$ \rm{keV} & $\gamma$(ground state) & $7.0\pm3.0\times10^{-2}$ \rm eV\\
 (Included) & & & 1 & & & n (elastic) $\approx$100\%& $33\pm6$ \rm{keV}\\
\hline
\be{8}, $2^{+}$, 16.922\ \rm MeV & $\li{7} + p$ & 1 & 2 & -333.1 \rm keV & $74.0\pm0.4$\ \rm{keV}& $\gamma$(ground state) & $8.4\pm1.4\times10^{-2}$ \rm eV\\
& & & 1 & & & $\gamma (3.04\ \rm MeV)$ &$<2.80\pm0.18$ \rm eV\\
& & & 2 & & & $\alpha\approx$ 100\% & $\approx$\ 74.0\ \rm{keV} \\
& & & 1 & & & p (elastic) & \rm unknown\\
\hline
\be{8}, $1^{+}$, 17.640\ \rm MeV & $\li{7} + p$ & 1 & 1 & 384.9 $\rm{keV}$ & 10.7 \rm{keV} & $\gamma$(ground state) & 16.7 \rm eV\\
& & & 1 & & & $\gamma (3.04\ \rm MeV)$ &$6.7\pm1.3$ \rm eV\\
& & & 2 & & & $\gamma (3.04\ \rm MeV)$ &$0.12\pm0.05$ \rm eV\\
& & & 1 & & & $\gamma (16.63\ \rm MeV)$ &$(3.2\pm0.3)\times10^{-2}$ \rm eV\\
& & & 1 & & & $\gamma (16.92\ \rm MeV)$ &$(1.3\pm0.3)\times10^{-3}$ \rm eV\\
& & & 1 & & & p (elastic) 98.8\%& 10.57\ \rm keV \\
\hline
\be{8}, $2^{-}$, 18.91\ \rm MeV & $\be{7} + n$ & 0 & 1 & 10.3 $\rm{keV}$ & 122 \rm{keV*} & $\gamma$(16.922 \rm MeV) & $9.9\pm4.3\times10^{-2}$ \rm eV\\
  (Included) & & & 1 & & & $\gamma$ (16.626 \rm MeV) &$0.17\pm0.07$ \rm eV\\
& & & 0 & & & p & $<105.1$\ \rm keV*\\
& & & 2 & & & $p + \li7^{*}$ (0.4776 MeV) & $<105.1$\ \rm keV*\\
& & & 0 & & & n (elastic) & 16.65 \rm keV*\\
\hline
\be{8}, $3^{+}$, 19.07\ \rm MeV & $\be{7} + n$ & 1 & 1 & 170.3 $\rm{keV}$ & $270\pm20\ \rm{keV}$ & p $\approx$ 100\% & $< 270$ \rm{keV} \\
  (Included) & & & 3 & & & $p + \li7^{*}$ (0.4776 MeV) & $< 270$ \rm{keV} \\
& & & 1 & & & $\gamma (3.03\ \rm MeV)$ &10.5 \rm eV\\
& & & 1 & & & n (elastic) &\rm unknown\\
\hline
\be{8}, $3^{+}$, 19.235\ \rm MeV& $\be{7} + n$ & 1 & 1 & 335.3 $\rm{keV}$ & $227\pm16\ \rm{keV}$ & p $\approx$\ 50\%& \ $\approx$\ 113.5 \rm{keV} \\
  (Included) & & & 1 & & & $\gamma (3.03\ \rm MeV)$ &10.5 \rm eV\\
& & & 1 & & & n (elastic) $\approx$\ 50\% &$\approx$\ 113.5 \rm keV \\
\hline
\be{8}, $1^{-}$, 19.40\ \rm MeV & $\be{7} + n$ & 0 & 0 & 500.3 $\rm{keV}$ & $645\ \rm{keV}$ & p & \rm unknown\\
& & & 0 & & & $p + \li7^{*}$ (0.4776 MeV) & \rm unknown\\
& & & 0 & & & n (elastic) &\rm unknown\\
& & & 1 & & & $\alpha$ & \rm unknown\\
\hline
$\bor{8}^{\rm g.s.}$, $2^{+}$, 0 \rm MeV & $\be{7} + p$ & 1 & 1 & -0.1375 MeV &  unknown &  $p$ (elastic) & unknown \\
  & & & 0 & & & EC$\rightarrow \be8$ & $8.5\times 10^{-19}\ \rm eV$ \\
\hline
\bor{8}, $1^{+}$, 0.7695\ \rm MeV & $\be{7} + p$ & 1 & 1 & 630\ $\pm3\ \rm{keV}$ & $35.7\pm0.6\  \rm{keV}$ & $\gamma$ (ground state) & $25.2\pm1.1$ \rm{meV}\\
  (Included) & & & 1 & & & p (elastic) 100\% & $35.7\pm0.6\ \rm{keV}$\\
\hline
\hline
\end{tabular}
}
}
\end{center}
\caption{This table lists the potential resonances in \li8, \be8 and \bor8 which may achieve required destruction of mass 7. These are all allowed by selection rules and includes some resonances already accounted for in determining the current theoretical \li7 abundance indicated as (Included). The entrance and exit channels along with their partial and total widths ($\gtot$), minimum angular momenta ($L_{\rm init},L_{\rm fin}$) as well as resonance energies are listed wherever experimental data are available. The starred widths are a result of fits from R-matrix analysis. The list includes final products in ground and excited states with the latter marked with a star in the superscript.}
\label{table:candidatesbebor8}
\end{table}

As seen in Table~\ref{table:candidatesbebor8},
the only resonance energy level of interest in the \li8 
compound nucleus at 2.255 MeV is already accounted 
for in the $\li7 + n$  reaction. In the $\be{8}$ compound nucleus, 
there are six levels of 
relevance for destroying either $\li{7}$ or \be7 at 16.922, 17.64, 18.91, 19.07, 19.24 
and 19.40 MeV within our limit on $\eres$. 
The 16.922 MeV level is 
more than 300 \rm keV below threshold and has 
a maximum total width of only 74 keV. 
Therefore, it is expected to have a weak effect. 
The 17.64 MeV level has 
typically low photon widths ($\approx 20 \ \rm eV$ ) 
and a total width of 10.7 $\rm keV$. But this state's decay is dominated 
by the elastic channel which makes this channel uninteresting. 

The energy level diagram 
for $\be{8}$ \cite{tunlbe8} shows the initial state, 
$\be{7}+n$ at an entrance energy of $E = 18.8997$ MeV bringing the
18.91, 19.07, 19.235 and 19.40 MeV levels into play. From 
among these the effect of the 18.91, 19.07 and 19.235 MeV 
resonances are already accounted for in the 
well known $\be7(n,p)\li7$ reaction \cite{Cyburt:2004cq}. 
The (18.91 MeV, $2^{-}$), resonance with $L_{\rm init} = 0$  
is the dominant contributor \cite{Coc:2003ce,Adahchour:2003}. Being 
a broad resonance with a total width of $\approx$ 122 keV, 
the Breit-Wigner form is not used and instead an R-matrix fit 
to the data \cite{Cyburt:2004cq}, is 
used to evaluate the contribution of the resonant rate. 
The remaining level at 19.40  
should also contribute 
to this reaction through ground and excited states. 
Only the 19.40 MeV channel can 
have an $\alpha$ exit channel due to parity 
considerations.  
And this resonance, despite a high resonance energy of 
$\approx 500$ keV, can in principle be 
important due to its large total width of 645 keV, if 
the proton branching ratio is high.

\begin{figure}[htbp]
\begin{center}
\rotatebox{90}{
\scalebox{0.8}{
\includegraphics[width=1.0\textwidth]{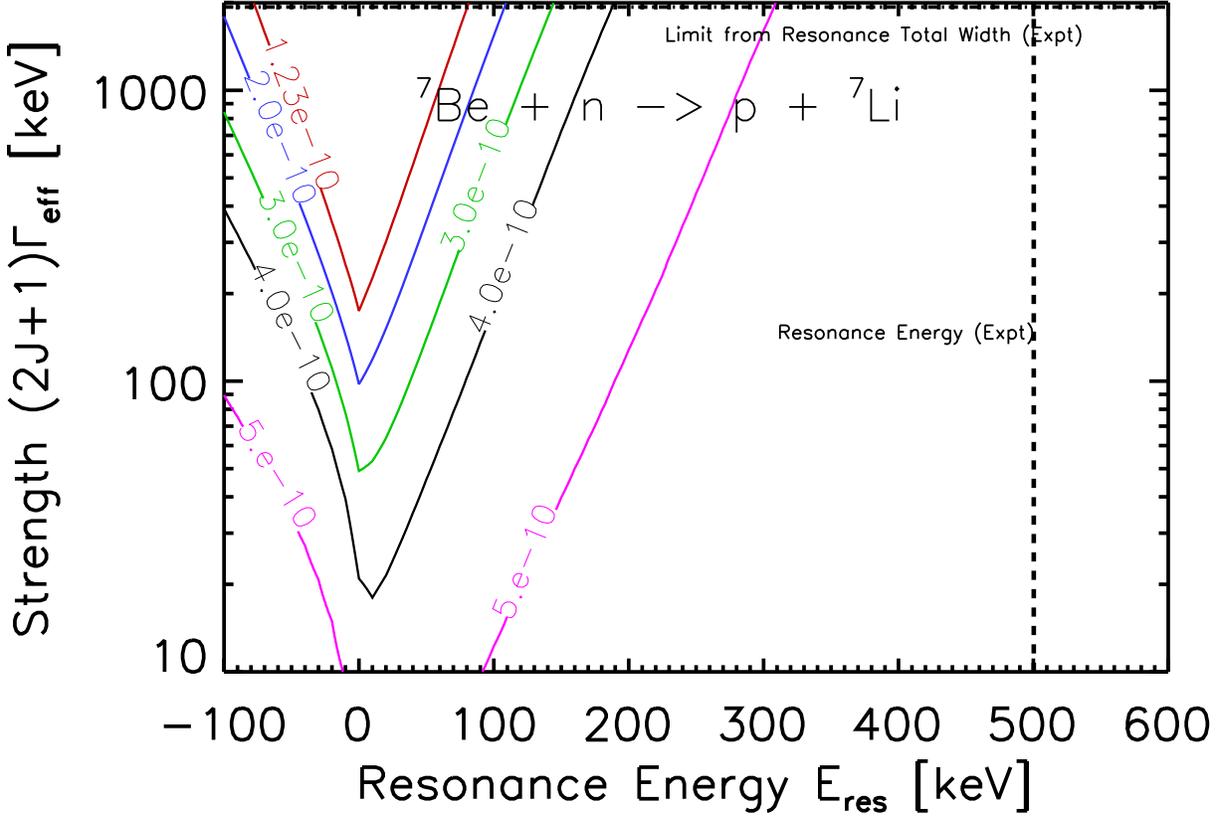}
}
}
\caption{The effect of resonances in the \be{8} compound nucleus involving initial states \be{7}$+n$. It shows the range of values for the
product of the resonant state spin degeneracy and resonance strength, 
$(2J+1)\geff$, versus 
the resonance energy.
Contours indicate where the lithium abundance is 
reduced to ${\rm \li7/H} = 1.23\times10^{-10}$ (red), $2.0\times10^{-10}$ (blue), $3.0\times10^{-10}$ (green) $4.0\times10^{-10}$ (black) and $5.0\times10^{-10}$ (magenta). 
Normal resonances have $\eres>0$, while 
subthreshold resonances lie in the $\eres<0$.
The horizontal dot-dashed black line is the experimental value of the strength of the resonance corresponding to the 19.40 MeV energy level.  The vertical dashed black line shows the position 
of $\eres$ for the same state.
}
\label{fig:8Becontours}
\end{center}
\end{figure}

Figure~\ref{fig:8Becontours} shows the \li7 abundance in the
($\Gamma_{\rm eff}, E_{\rm res}$) plane for 
the $\be7(n,p)\li7$ reaction.
Contours for 
\li7/H~$\times 10^{10}$ = 1.23, 2.0, 3.0, 4.0, and 5.0 (as labelled)
are plotted as functions of the effective width and resonant energy. 
Below $\approx\ (2J_* + 1) 40 =120 \ \rm keV$, we expect our results based on 
the narrow resonance approximation to be quite accurate.
As one can see from this figure, to bring the \li7 abundance down close to 
observed values, one would require a very low resonance energy (of order
$\pm 30$ keV) with a relatively large effective width.  
Unfortunately, the 19.40 MeV level of \be8 corresponds to $E_{\rm res} = 500$ keV
as shown by the vertical dashed line and does not make any real impact on the
\li7 abundance.

\begin{figure}[htbp]
\begin{center}
\rotatebox{90}{
\scalebox{0.8}{
\includegraphics[width=1.0\textwidth]{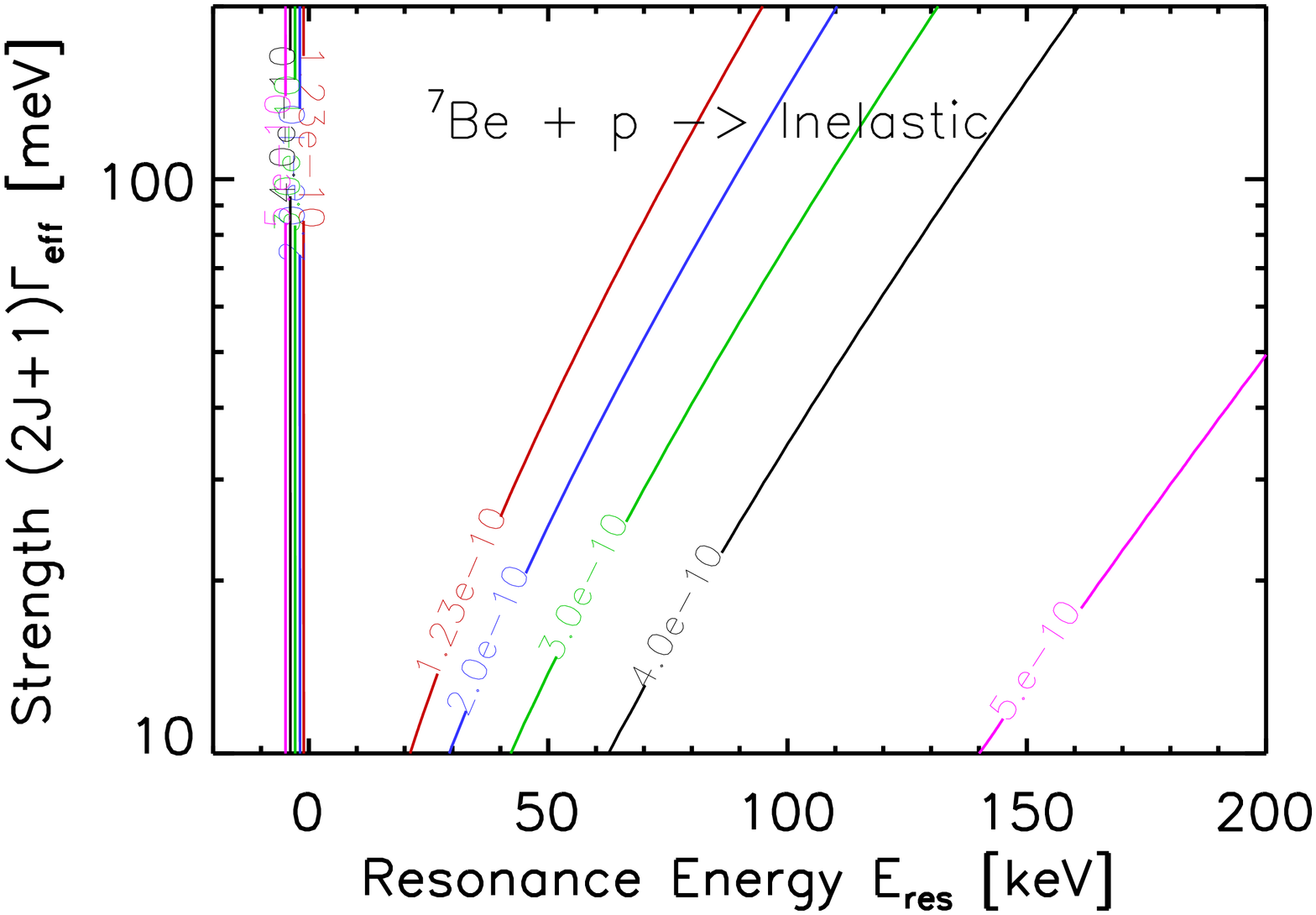}
}
}
\caption{As in Fig. \ref{fig:8Becontours} for the resonances in the \bor{8} compound nucleus involving initial states \be{7}$+p$.   }
\label{fig:8Bcontours}
\end{center}
\end{figure}

Figure \ref{fig:8Bcontours}
shows the effect of a \bor8 resonance with \be{7} and $p$ 
in the initial state,
plotted in the ($\Gamma_{\rm eff}, E_{\rm res}$) plane  with contours of 
constant mass 7 abundances. 
According to Fig.~\ref{fig:8Bcontours} for resonance energies of a few 10's
of \rk 's, resonance strength of a few meV is sufficient 
to attain the observational value of mass 7. However, from 
the energy level diagram for $\bor{8}$, \cite{tunlbor8}, 
the closest resonant energy level, $E_{\rm ex}$ 
is at 0.7695 MeV \cite{tunlbor8}, 
whose effect is already included via the 
$\be7(p,\gamma)\bor8$ reaction.
The experimental value of resonance energy is 632 \rm keV which is off the scale in this figure.
The only other close energy level to the 
$\be{7}+p$ entrance channel is at -0.1375 MeV 
which means that the ground state is a sub-threshold state. 
This is not the usual resonant reaction, since 
the ground state doesn't have a width in the sense 
we refer to a width for the other reactions. But at 
these energies, the astrophysical $S$-factor is 
$\approx 10 $ eV-barn which is very small and will yield a low cross-section. 
This too is off scale in the figure and
verifies that the $\be7(p,\gamma)\bor8$ reaction does not yield an important destruction channel.

\subsection{\label{compnuc9} A = 9 Compound Nucleus}

\begin{figure}[htbp]
\begin{center}
\rotatebox{90}{
\scalebox{0.8}{
\includegraphics[width=1.0\textwidth]{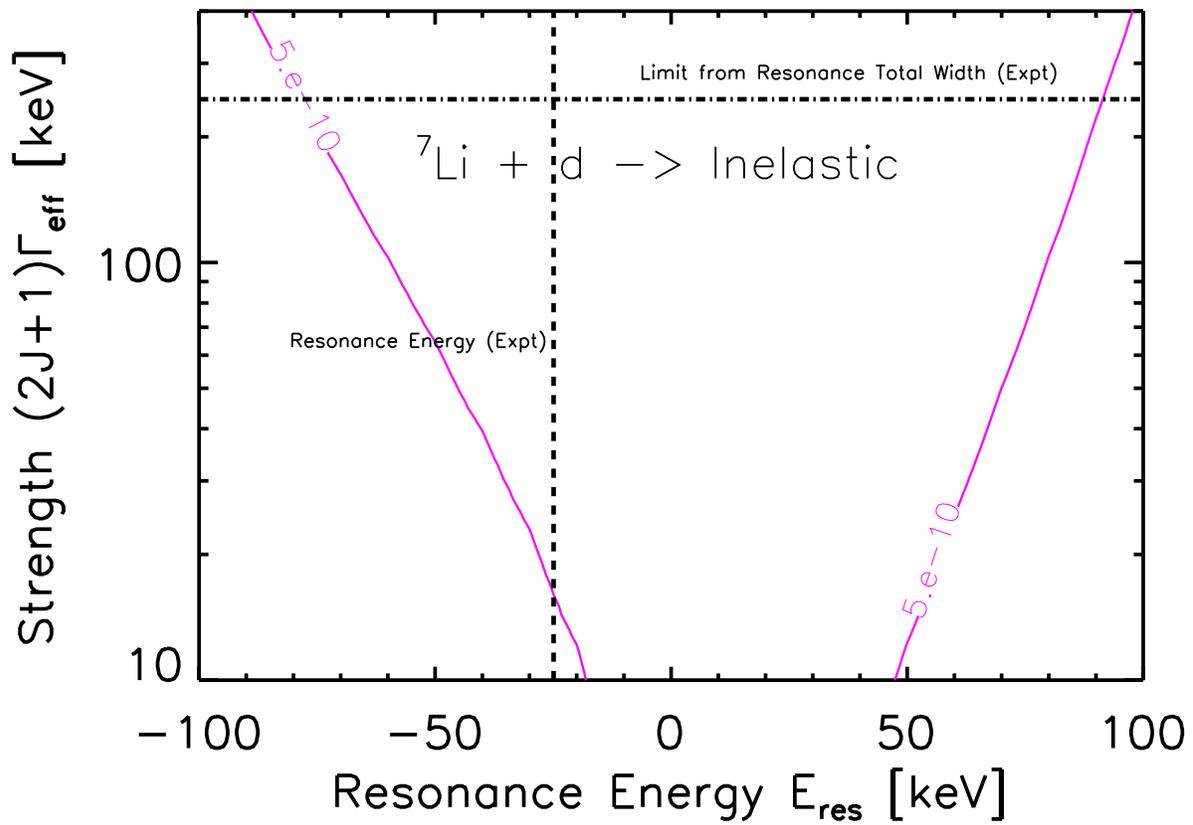}
}
}
\caption{As in Fig. \ref{fig:8Becontours} for the resonances
in the \be9 compound nucleus. 
 }
\label{fig:7lidpcontours}
\end{center}
\end{figure}

The energy level diagram for $\be{9}$, \cite{tunlbe9} 
shows energy levels of interest at 16.671, 16.9752 
and 17.298 MeV; these appear in Table \ref{table:candidatesbe9}. 
The $\li{7}+d$ entrance channel sits at 
16.6959 MeV. The lowest lying resonant state is at 16.671
MeV and is a sub-threshold state with $E_{\rm res} = -24.9$ keV
which lies within the total width of 41 keV.
This resonance is thus obviously tantalizing--it is well-tuned energetically
and involves an abundant, stable projectile.
The \li7 abundance contours for the \be9 resonance states are
shown in Fig.~\ref{fig:7lidpcontours}.
Perhaps disappointingly, the figure shows
that the effect on primordial mass 7 is minor.
This illustrates the inability of
direct \li7 
destruction channels to reduce the mass-7 abundance,
as explained in \S \ref{semi-analytical}.
Given the overall difficulty of this channel, 
it is clear that the
other possible resonant energy levels (16.9752 MeV and 17.298 MeV) 
also fail to
substantially reduce the mass-7 abundance.

\begin{table}[ht!]
\begin{center}
\rotatebox{0}{
\scalebox{0.76}{
\begin{tabular}{|c|c|c|c|c|c|c|c|}
\hline
\hline
Compound Nucleus, & Initial & $L_{\rm init}$ & $L_{\rm fin}$ & $E_{\rm res}$ & $\gtot$ & \rm Exit & \rm Exit Channel \\ 
$J^\pi, E_{\rm ex}$ & State & & & & & Channels & Width \\
\hline
\be{9},  & $\li{7} + d$ & 1 & \rm unknown & -24.9 $\rm{keV}$ & $41\pm4$\ \rm{keV} &$\gamma$ & \rm unknown\\
$(5/2^{+})$, 16.671 \rm MeV & & & 2 & & & $n + \be8$ & \rm unknown\\
& & & 0 & & & $n + \be8^{*}$ (3.03 MeV) & \rm unknown\\
& & & 2 & & & $n + \be8^{*}$ (11.35 MeV) & \rm unknown\\
& & & 0 & & & p & \rm unknown\\
& & & 1 & & & $\alpha$ & \rm unknown\\
& & & 1 & & & d (elastic) & \rm unknown\\
\hline
\be{9}, & $\li{7} + d$ & 0 & 1 & 279.3 $\rm{keV}$ & $389\pm10$\ \rm{eV} &$\gamma$ (ground state) & $16.9\pm1.0$ \rm eV\\
 $1/2^{-}$, 16.9752 \rm MeV & & & 1 & & & $\gamma$ (1.68\ \rm MeV) & $1.99\pm0.15$ \rm eV\\
& & & 2 & & & $\gamma$ (2.43\ \rm MeV) & $0.56\pm0.12$ \rm eV\\
& & & 1 & & & $\gamma$ (2.78\ \rm MeV) & $2.2\pm0.7$ \rm eV\\
& & & \rm unknown & & & $\gamma$ (Unknown level TUNL) & $< 0.8$ \rm eV\\
& & & 1 & & & $\gamma$ (4.70\ \rm MeV) & $2.2\pm0.3$ \rm eV\\
& & & 1 & & & p & $12^{+12}_{-6}$ \rm eV\\
& & & 1 & & & n & $< 288$ \rm eV \\
& & & 1 & & & $n + \be8^{*}$ (3.03 MeV) & $< 288$ \rm eV\\
& & & 3 & & & $n + \be8^{*}$ (11.35 MeV) & $< 288$ \rm eV\\
& & & 2 & & & $\alpha$ & $< 241$ \rm eV \\
& & & 0 & & & d (elastic) & $62\pm10$ \rm eV\\
\hline     
\be{9}, & $\li{7} + d$ & 0 & \rm unknown & 602.1 $\rm{keV}$ & 200\ \rm{keV} &$\gamma$ (ground state) & \rm unknown\\
  $(5/2)^{-}$, 17.298 \rm MeV & & & 1 & & & p & 194.4\ \rm keV \\
(Included) & & & 3 & & & $n + \be8$ & \rm unknown \\
& & & 1 & & & $n + \be8^{*}$ (3.03 MeV) & \rm unknown\\
& & & 1 & & & $n + \be8^{*}$ (11.35 MeV) & \rm unknown\\
& & & 2 & & & $\alpha$ &\rm unknown\\
& & & 0 & & & d (elastic)& \rm unknown\\
\hline
\hline
\end{tabular}
}
}
\end{center}
\caption{As in Table \ref{table:candidatesbebor8} listing the potential resonances in \be9.}
\label{table:candidatesbe9}
\end{table}

\begin{table}[ht!]
\begin{center}
\rotatebox{0}{
\scalebox{0.87}{
\begin{tabular}{|c|c|c|c|c|c|c|c|}
\hline
\hline
Compound Nucleus, & Initial & $L_{\rm init}$ & $L_{\rm fin}$ & $E_{\rm res}$ & $\gtot$ & \rm Exit & \rm Exit Channel \\ 
$J^\pi, E_{\rm ex}$ & State & & & & & Channels & Width \\
\hline
\bor{9}, $(5/2^{+})$, 16.71\ \rm MeV & $\be{7} + d$ & 1 & 2 & 219.9 $\rm{keV}$ & \rm unknown & $p + \be{8}$ &\rm unknown\\
& & & 0 & & & $p + \be8^{*}$ (3.03 MeV)&\rm unknown\\
& & & 2 & & & $p + \be8^{*}$ (11.35 MeV)&\rm unknown\\
& & & 0 & & & $p + \be8^{*}$ (16.626 MeV)&\rm unknown\\
& & & 0 & & & $p + \be8^{*}$ (16.922 MeV)&\rm unknown\\
& & & 2 & & & \he3 &\rm unknown\\
& & & 1 & & & $\alpha + \li5$ &\rm unknown\\
& & & 3 & & & $\alpha + \li5^{*}$ (1.49 MeV) &\rm unknown\\
& & & 1 & & & d (elastic)& \rm unknown\\
\hline
\bor{9}, $(1/2^{-})$, 17.076\ \rm MeV & $\be{7} + d$ & 0 & 1 & 585.9 $\rm{keV}$ & 22 \rm keV & $p + \be{8}$ &\rm unknown\\
& & & 1 & & & $p + \be8^{*}$ (3.03 MeV)&\rm unknown\\
& & & 3 & & & $p + \be8^{*}$ (11.35 MeV)&\rm unknown\\
& & & 1 & & & $p + \be8^{*}$ (16.626 MeV)&\rm unknown\\
& & & 1 & & & \he3 &\rm unknown\\
& & & 2 & & & $\alpha + \li5$ &\rm unknown\\
& & & 0 & & & $\alpha + \li5^{*}$ (1.49 MeV) &\rm unknown\\
& & & 0 & & & d (elastic)& \rm unknown\\
\hline
\hline
\hline
\end{tabular}
}
}
\end{center}
\caption{As in Table \ref{table:candidatesbebor8} listing the potential resonances in 
\bor9.} 
\label{table:candidatesbor9}
\end{table}

\begin{figure}[htbp]
\begin{center}
\rotatebox{90}{
\scalebox{0.8}{
\includegraphics[width=1.0\textwidth]{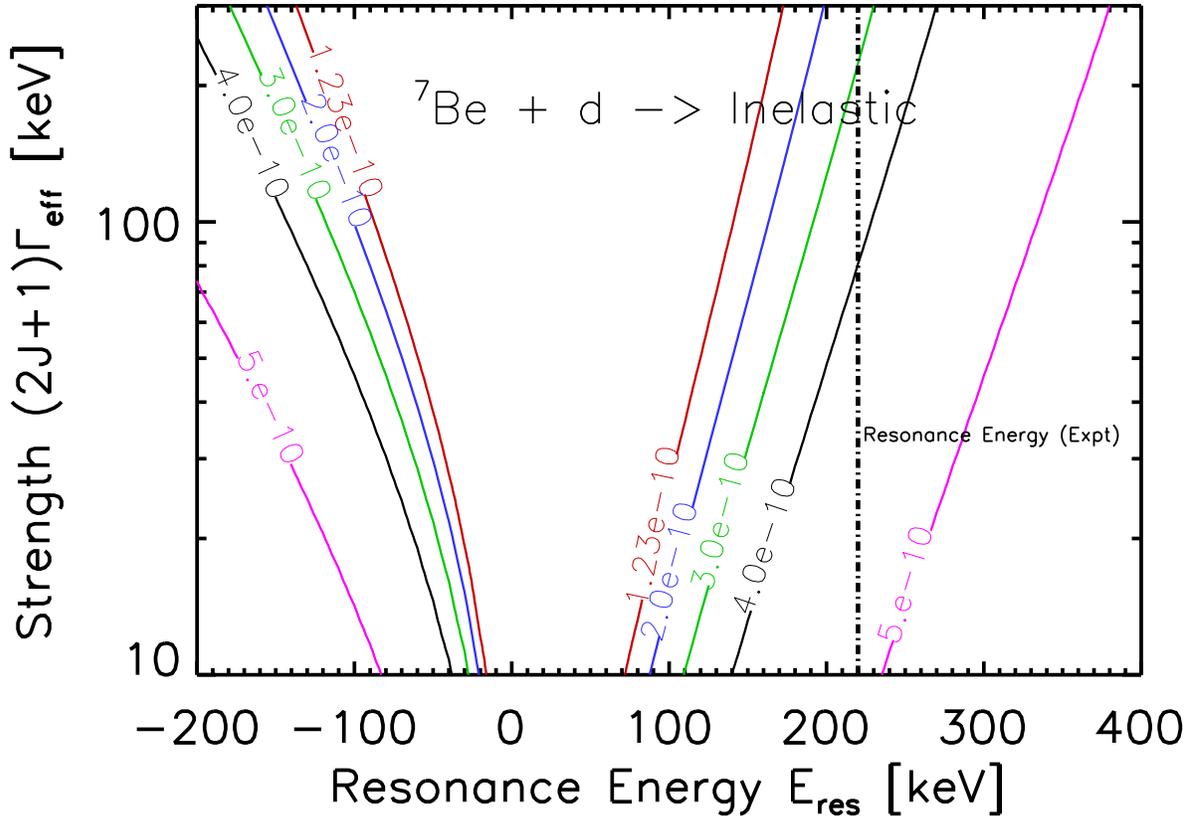}
}
}
\caption{As in Fig. \ref{fig:8Becontours} for the resonances
in the \bor9 compound nucleus.
The vertical dashed 
line at 220 keV indicates the experimental central 
value of the resonance energy of the 16.71 MeV level. 
 }
\label{fig:7bedpcontours}
\end{center}
\end{figure}

The $\bor{9}$ compound nucleus is relevant for 
studying the effect of the $\be{7}(d,p)2 \alpha$ and its 
competitors such as $\be{7}(d,\he{3})\li{6}$ and 
$\be{7}(d,\alpha)\li{5}$.
As seen in Table~\ref{table:candidatesbor9}, the only two levels 
of interest here are the 16.71 and 17.076 MeV levels. 
The 16.71 MeV level corresponds 
to a resonance energy of 220 \rm{keV} as shown by the 
vertical dashed line in Fig.~\ref{fig:7bedpcontours}
 \cite{tunlbor9}. The widths are unknown experimentally. 
The approximate narrow resonance limit on the 
 resonance width which is shown 
 by the horizontal solid line is around 40\ $\rm keV$. 
The $p$ exit 
channel leads to the $\be7(d,p)\be8^{*}$ reaction through 
the excited state at 16.63 MeV in \be8. 
This should eventually lead to formation of alpha particles.
Fig.~\ref{fig:7bedpcontours} shows the effect of the 
16.71 MeV resonance on the mass-7 abundance as 
a function of the resonance strength and energy under the 
narrow resonance approximation. From the plot, 
we see that the \li7 abundance is reduced by  
50\% for $(2J+1)\geff = 240\ \rm keV$. 
This state has $J= 5/2$ and therefore, 
a value $\geff = 40\ \rm keV $ or more 
will have substantial impact on the problem. 
Furthermore, 
as $\Gamma_{L} \ge \geff$, we require 
$\Gamma_{L} \ga 40\ \rm keV $. This result 
confirms 
the conclusion of \cite{Cyburt:2009cf}.
Later in \S~\ref{reducedlist} we will see 
how this compares with theoretical limits.
As the decay widths are largely unknown, experimental 
data on the width are needed. 

The state at 17.076 MeV corresponds to a resonant energy of $ E_{\rm res} = 586$ keV
and is beyond the scale shown in Fig.~\ref{fig:7bedpcontours}.  
A solution using this state is very unlikely.

\subsection{\label{compnuc10} A = 10 Compound Nucleus}

\begin{figure}[htbp]
\begin{center}
\mbox{
\subfigure{\includegraphics[width=0.35\textwidth,angle=90]{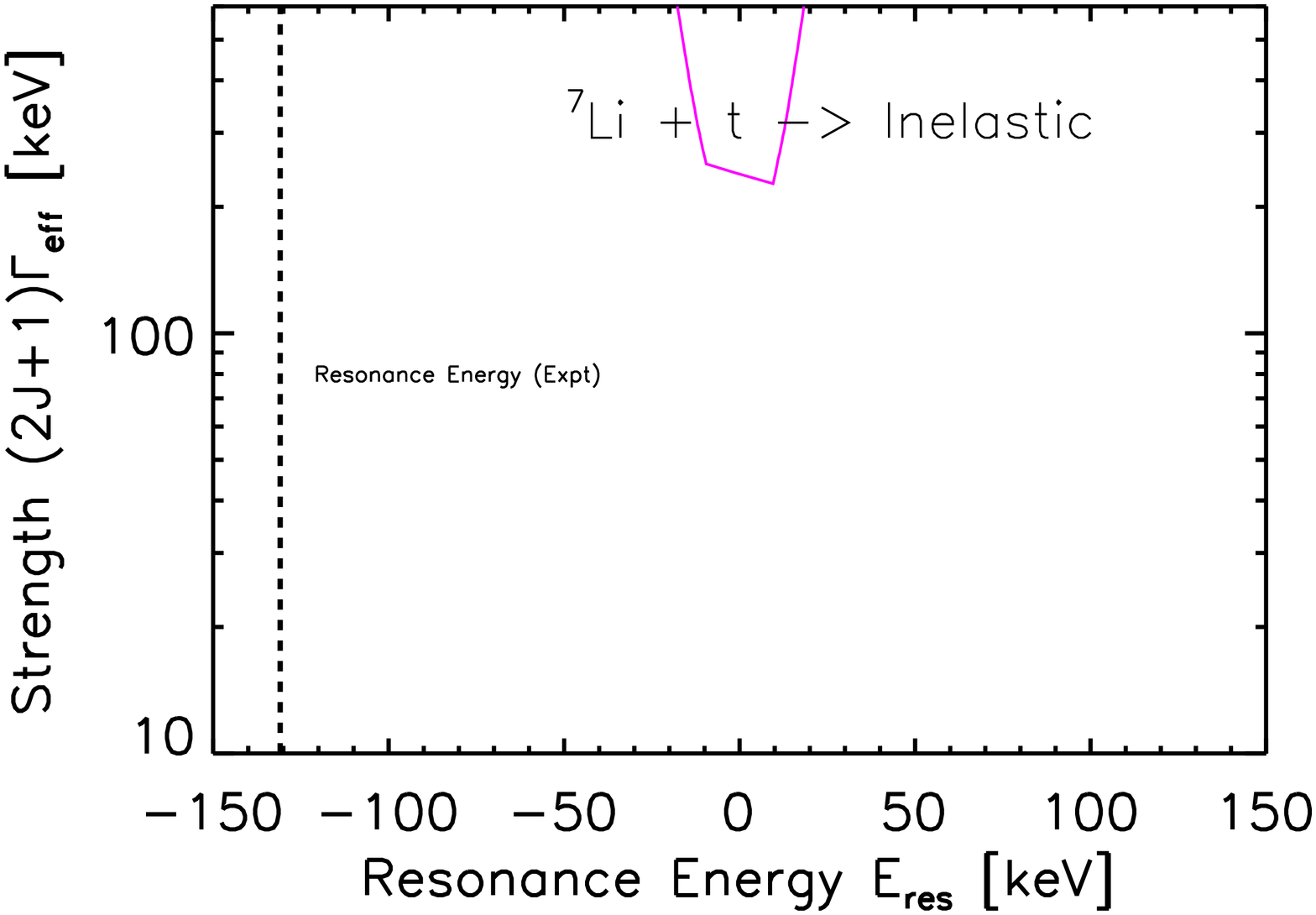}}
\subfigure{\includegraphics[width=0.35\textwidth,angle=90]{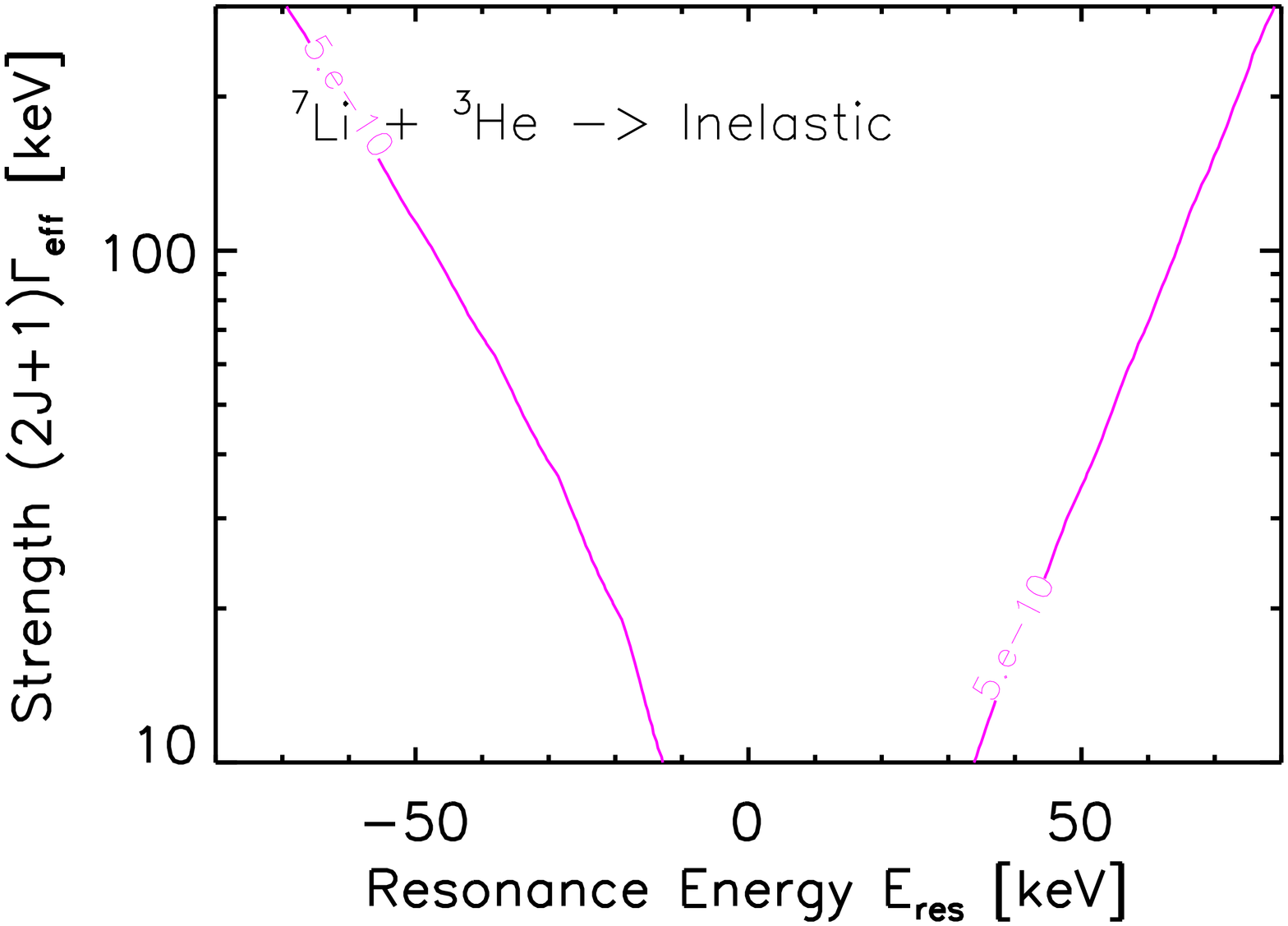}}}
\caption{As in Fig. \ref{fig:8Becontours} for the resonances in the \be{10} compound nucleus involving the initial state \li7$+t$ ({\em left}),
and in the \bor{10} compound nucleus involving the initial state \li7$+\he3$.
({\em right}). }
\label{fig:7A3Hecontours}
\end{center}
\end{figure}

Table \ref{table:candidatesbe10} shows that
the $\be{10}$ compound nucleus has energy levels at 
17.12 and 17.79 MeV \cite{tunlbe10} which are close to 
the initial state $\li{7}+t$ at 17.2509 MeV.  The former 
is far below threshold and does not contribute to \li7 destruction.
The 17.79 MeV level is 
around 540 $\rm{keV}$ above the entrance energy and its
spin and parity are unknown. The total width 
\cite{tunlbe10} is $\gtot = 112$ keV which implies a small 
overlap with the entry channel which 
renders this resonance insignificant 
despite having a number of $n$ exit channels with both 
ground state and excited states of \be9. 
As seen in Fig.~\ref{fig:7A3Hecontours}, 
the effect of $\li7+t$ is small for the interesting region of parameter space.

\begin{table}[h!]
\begin{center}
\rotatebox{0}{
\scalebox{0.76}{
\begin{tabular}{|c|c|c|c|c|c|c|c|}
\hline
\hline
Compound Nucleus,& Initial State & $L_{\rm init}$ & $L_{\rm fin}$ & $E_{\rm res}$ & $\gtot$ & \rm Exit & \rm Exit Channel \\ 
$J^\pi, E_{\rm ex}$& & & & & &Channels  & Width \\
\hline
\be{10},  & $\li{7} + t$ & 0 & 0 & -130.9 $\rm{keV}$ & $\approx 150\ \rm{keV}$ & $n+\be9$ & \rm unknown\\
$(2^{-})$, 17.12 MeV& & & 1 & & & $n+\be9^{*}$ (1.684 MeV)& \rm unknown\\
& & & 0 & & & $n+\be9^{*}$ (2.4294 MeV)& \rm unknown\\
& & & 2 & & & $n+\be9^{*}$ (2.78 MeV)& \rm unknown\\
& & & 1 & & & $n+\be9^{*}$ (3.049 MeV)& \rm unknown\\
& & & 1 & & & $n+\be9^{*}$ (4.704 MeV)& \rm unknown\\
& & & 0 & & & $n+\be9^{*}$ (5.59 MeV)& \rm unknown\\
& & & 2 & & & $n+\be9^{*}$ (6.38 MeV)& \rm unknown\\
& & & 3 & & & $n+\be9^{*}$ (6.76 MeV)& \rm unknown\\
& & & 0 & & & $n+\be9^{*}$ (7.94 MeV)& \rm unknown\\
& & & 0 & & & t (elastic) & \rm unknown\\
\hline
\be{10}, & $\li{7} + t$ & \rm unknown & \rm unknown & 539.1 $\rm{keV}$ & $112\pm35\ \rm{keV}$ & $\gamma$ & $3+2$ \rm eV\\
\rm unknown, 17.79 MeV & & & \rm unknown & & & $n + \be9$ & $<$\ 77\ \rm{keV} \\
& & & \rm unknown & & & $n+\be9^{*}$ (1.684 MeV)& $<$\ 77\ \rm{keV}\\
& & & \rm unknown & & & $n+\be9^{*}$ (2.4294 MeV)& $<$\ 77\ \rm{keV}\\
& & & \rm unknown & & & $n+\be9^{*}$ (2.78 MeV)& $<$\ 77\ \rm{keV} \\
& & & \rm unknown & & & $n+\be9^{*}$ (3.049 MeV)& $<$\ 77\ \rm{keV} \\
& & &\rm unknown & & & $n+\be9^{*}$ (4.704 MeV)& $<$\ 77\ \rm{keV} \\
& & & \rm unknown & & & $n+\be9^{*}$ (5.59 MeV)& $<$\ 77\ \rm{keV} \\
& & & \rm unknown & & & $n+\be9^{*}$ (6.38 MeV)& $<$\ 77\ \rm{keV} \\
& & & \rm unknown & & & $n+\be9^{*}$ (6.76 MeV)& $<$\ 77\ \rm{keV} \\
& & & \rm unknown & & & $n+\be9^{*}$ (7.94 MeV)& $<$\ 77\ \rm{keV} \\
& & & \rm unknown & & & t (elastic) & 78\ \rm{keV}\\
\hline
\end{tabular}
}
}
\end{center}
\caption{As in Table \ref{table:candidatesbebor8} listing the potential resonant reactions in \be{10}. } 
\label{table:candidatesbe10}
\end{table}

\begin{table}[ht!]
\begin{center}
\rotatebox{0}{
\scalebox{0.77}{
\begin{tabular}{|c|c|c|c|c|c|c|c|}
\hline
\hline
Compound Nucleus, & Initial & $L_{\rm init}$ & $L_{\rm fin}$ & $E_{\rm res}$ & $\gtot$ & \rm Exit & \rm Exit Channel \\ 
$J^\pi, E_{\rm ex}$& State & & & & &  Channels & Width \\
\hline
\bor{10},  & $\li{7} + \he3$ & \rm unknown & \rm unknown & 411.7 $\rm{keV}$ & $1500\pm300\ \rm{keV}$ & $p+\be9$ &\rm unknown\\
\rm unknown, (18.2 MeV) & & & \rm unknown & & & $p+\be9^{*}$ (1.684 MeV)& \rm unknown\\
& & & \rm unknown & & & $p+\be9^{*}$ (2.4294 MeV)& \rm unknown\\
& & & \rm unknown & & & $p+\be9^{*}$ (2.78 MeV)& \rm unknown\\
& & & \rm unknown & & & $p+\be9^{*}$ (3.049 MeV)& \rm unknown\\
& & & \rm unknown & & & $p+\be9^{*}$ (4.704 MeV)& \rm unknown\\
& & & \rm unknown & & & $p+\be9^{*}$ (5.59 MeV)& \rm unknown\\
& & & \rm unknown & & & $p+\be9^{*}$ (6.38 MeV)& \rm unknown\\
& & & \rm unknown & & & $p+\be9^{*}$ (6.76 MeV)& \rm unknown\\
& & & \rm unknown & & & $p+\be9^{*}$ (7.94 MeV)& \rm unknown\\
& & & \rm unknown & & & $p+\be9^{*}$ (11.283 MeV)& \rm unknown\\
& & & \rm unknown & & & $n+\bor9$ & \rm unknown\\
& & & \rm unknown & & &  $n+\bor9^{*}$ (1.6 MeV) & \rm unknown\\
& & & \rm unknown & & &  $n+\bor9^{*}$ (2.361 MeV) & \rm unknown\\
& & & \rm unknown & & &  $n+\bor9^{*}$ (2.75 MeV) & \rm unknown\\
& & & \rm unknown & & &  $n+\bor9^{*}$ (2.788 MeV) & \rm unknown\\
& & & \rm unknown & & &  $n+\bor9^{*}$ (4.3 MeV) & \rm unknown\\
& & & \rm unknown & & &  $n+\bor9^{*}$ (6.97 MeV) & \rm unknown\\
& & & \rm unknown & & & $d + \be8$ &\rm unknown\\
& & & \rm unknown & & & $d + \be8^{*}$ (3.03 MeV)&\rm unknown\\
& & & \rm unknown & & & $d + \be8^{*}$ (11.35 MeV)&\rm unknown\\
& & & \rm unknown & & & t & \rm unknown\\
& & & \rm unknown & & & $\alpha + \li6$ & \rm unknown\\
& & & \rm unknown & & & $\alpha + \li6^{*}$ (2.186 MeV) & \rm unknown\\
& & & \rm unknown & & & $\alpha + \li6^{*}$ (3.563 MeV) & \rm unknown\\
& & & \rm unknown & & & $\alpha + \li6^{*}$ (4.31 MeV) & \rm unknown\\
& & & \rm unknown & & & $\alpha + \li6^{*}$ (5.37 MeV) & \rm unknown\\
& & & \rm unknown & & & \he3 (elastic) & \rm unknown\\
\hline
\hline
\end{tabular}
}
}
\end{center}
\caption{As in Table \ref{table:candidatesbebor8} listing the ground and excited final 
states for the 18.2 MeV energy level in \bor{10}. } 
\label{table:candidatesbor10a}
\end{table}

\begin{table}[h!]
\begin{center}
\rotatebox{0}{
\scalebox{0.85}{
\begin{tabular}{|c|c|c|c|c|c|c|c|}
\hline
\hline
Compound Nucleus, & Initial & $L_{\rm init}$ & $L_{\rm fin}$ & $E_{\rm res}$ & $\gtot$ & \rm Exit & \rm Exit Channel \\ 
$J^\pi, E_{\rm ex}$& State & & & & &  Channels & Width \\
\hline
\bor{10},& $\li{7} + \he3$ & 0 & \rm unknown & 641.7 $\rm{keV}$ & $340\ \rm{keV}$ &$\gamma $(ground state) & $\ge 3$ \rm eV\\
 $2^{-}$, 18.43 MeV & & & \rm unknown & & & $\gamma$ (4.77 MeV) &$\ge 17 $\rm eV\\
& & & 0 & & & $n+\bor9$ & \rm unknown\\
& & & \rm unknown & & &  $n+\bor9^{*}$ (1.6 MeV) & \rm unknown\\
& & & 0 & & &  $n+\bor9^{*}$ (2.361 MeV) & \rm unknown\\
& & & 2 & & &  $n+\bor9^{*}$ (2.75 MeV) & \rm unknown\\
& & & 1 & & &  $n+\bor9^{*}$ (2.788 MeV) & \rm unknown\\
& & & \rm unknown & & &  $n+\bor9^{*}$ (4.3 MeV) & \rm unknown\\
& & & 2 & & &  $n+\bor9^{*}$ (6.97 MeV) & \rm unknown\\
& & & 0 & & & p + $\be9$& \rm unknown\\
& & & 1 & & & $p+\be9^{*}$ (1.684 MeV)& \rm unknown\\
& & & 0 & & & $p+\be9^{*}$ (2.4294 MeV)& \rm unknown\\
& & & 2 & & & $p+\be9^{*}$ (2.78 MeV)& \rm unknown\\
& & & 1 & & & $p+\be9^{*}$ (3.049 MeV)& \rm unknown\\
& & & 1 & & & $p+\be9^{*}$ (4.704 MeV)& \rm unknown\\
& & & 0 & & & $p+\be9^{*}$ (5.59 MeV)& \rm unknown\\
& & & 2 & & & $p+\be9^{*}$ (6.38 MeV)& \rm unknown\\
& & & 3 & & & $p+\be9^{*}$ (6.76 MeV)& \rm unknown\\
& & & 0 & & & $p+\be9^{*}$ (7.94 MeV)& \rm unknown\\
& & & 2 & & & $p+\be9^{*}$ (11.283 MeV)& \rm unknown\\
& & & 0 & & & $p + \be9^{*}$ (11.81 \rm MeV)& \rm unknown\\
& & & 1 & & & $d + \be8$ &\rm unknown\\
& & & 1 & & & $d + \be8^{*}$ (3.03 MeV)&\rm unknown\\
& & & 1 & & & $d + \be8^{*}$ (11.35 MeV)&\rm unknown\\
& & & 1 & & & $\alpha + \li6$ & \rm unknown\\
& & & 1 & & & $\alpha + \li6^{*}$ (2.186 MeV) & \rm unknown\\
& & & 1 & & & $\alpha + \li6^{*}$ (4.31 MeV) & \rm unknown\\
& & & 1 & & & $\alpha + \li6^{*}$ (5.37 MeV) & \rm unknown\\
& & & 1 & & & $\alpha + \li6^{*}$ (5.65 MeV) & \rm unknown\\
& & & 0 & & & $\he3$ (elastic) & \rm unknown\\
\hline
\hline
\end{tabular}
}
}
\end{center}
\caption{As in Table \ref{table:candidatesbebor8} listing the ground and excited final 
state channels for the 18.43 MeV energy level in \bor{10} 
for the $\li7+\he3$ initial state.}
\label{table:candidatesbor10b}
\end{table}

\begin{table}[h!]
\begin{center}
\rotatebox{0}{
\scalebox{0.85}{
\begin{tabular}{|c|c|c|c|c|c|c|c|}
\hline
\hline
Compound Nucleus, & Initial & $L_{\rm init}$ & $L_{\rm fin}$ & $E_{\rm res}$ & $\gtot$ & \rm Exit & \rm Exit Channel \\ 
$J^\pi, E_{\rm ex}$& State & & & & &  Channels & Width \\
\hline
\bor{10}, & $\be7 + t$ & 0 & \rm unknown & -239.1 $\rm{keV}$ & $340\ \rm{keV}$ &$\gamma $(ground state) & $\ge 3$ \rm eV\\
 $2^{-}$, 18.43 MeV & & & \rm unknown & & & $\gamma$ (4.77 MeV) &$\ge 17$ \rm eV\\
& & & 0 & & & $n+\bor9$ & \rm unknown\\
& & & \rm unknown & & &  $n+\bor9^{*}$ (1.6 MeV) & \rm unknown\\
& & & 0 & & &  $n+\bor9^{*}$ (2.361 MeV) & \rm unknown\\
& & & 2 & & &  $n+\bor9^{*}$ (2.75 MeV) & \rm unknown\\
& & & 1 & & &  $n+\bor9^{*}$ (2.788 MeV) & \rm unknown\\
& & & \rm unknown & & &  $n+\bor9^{*}$ (4.3 MeV) & \rm unknown\\
& & & 2 & & &  $n+\bor9^{*}$ (6.97 MeV) & \rm unknown\\
& & & 0 & & & p + $\be9$& \rm unknown\\
& & & 1 & & & $p+\be9^{*}$ (1.684 MeV)& \rm unknown\\
& & & 0 & & & $p+\be9^{*}$ (2.4294 MeV)& \rm unknown\\
& & & 2 & & & $p+\be9^{*}$ (2.78 MeV)& \rm unknown\\
& & & 1 & & & $p+\be9^{*}$ (3.049 MeV)& \rm unknown\\
& & & 1 & & & $p+\be9^{*}$ (4.704 MeV)& \rm unknown\\
& & & 0 & & & $p+\be9^{*}$ (5.59 MeV)& \rm unknown\\
& & & 2 & & & $p+\be9^{*}$ (6.38 MeV)& \rm unknown\\
& & & 3 & & & $p+\be9^{*}$ (6.76 MeV)& \rm unknown\\
& & & 0 & & & $p+\be9^{*}$ (7.94 MeV)& \rm unknown\\
& & & 2 & & & $p+\be9^{*}$ (11.283 MeV)& \rm unknown\\
& & & 0 & & & $p + \be9^{*}$ (11.81 \rm MeV)& \rm unknown\\
& & & 1 & & & $d + \be8$ &\rm unknown\\
& & & 1 & & & $d + \be8^{*}$ (3.03 MeV)&\rm unknown\\
& & & 1 & & & $d + \be8^{*}$ (11.35 MeV)&\rm unknown\\
& & & 1 & & & $\alpha + \li6$ & \rm unknown\\
& & & 1 & & & $\alpha + \li6^{*}$ ( 2.186 MeV) & \rm unknown\\
& & & 1 & & & $\alpha + \li6^{*}$ ( 4.31 MeV) & \rm unknown\\
& & & 1 & & & $\alpha + \li6^{*}$ ( 5.37 MeV) & \rm unknown\\
& & & 1 & & & $\alpha + \li6^{*}$ ( 5.65 MeV) & \rm unknown\\
& & & 0 & & & t (elastic) & \rm unknown\\
\hline
\hline
\end{tabular}
}
}
\end{center}
\caption{As in Table \ref{table:candidatesbebor8} listing the ground and excited final 
state channels for the 18.43 MeV energy level in \bor{10} 
for the $\be7+t$ initial state.}
\label{table:candidatesbor10c}
\end{table}

\begin{table}[h!]
\begin{center}
\rotatebox{0}{
\scalebox{0.79}{
\begin{tabular}{|c|c|c|c|c|c|c|c|}
\hline
\hline
Compound Nucleus, & Initial State & $L_{\rm init}$ & $L_{\rm fin}$ & $E_{\rm res}$ & $\gtot$ & \rm Exit & \rm Exit Channel \\ 
$J^\pi, E_{\rm ex}$& & & & & &  Channels & Width \\
\hline
\bor{10},  & $\be{7} + t$ & 1 & \rm unknown & 130.9 $\rm{keV}$ & $< 600\ \rm{keV}$ &$\gamma$\ (0.72 \rm MeV)&$\ge 20$ \rm eV\\
$2^{+}$, 18.80 MeV & & & \rm unknown & & & $\gamma$ (3.59 MeV) &$\ge 20$ \rm eV\\
& & & 1 & & & $n+\bor9$ & \rm unknown\\
& & & \rm unknown & & &  $n+\bor9^{*}$ (1.6 MeV) & \rm unknown\\
& & & 1 & & &  $n+\bor9^{*}$ (2.361 MeV) & \rm unknown\\
& & & 1 & & &  $n+\bor9^{*}$ (2.75 MeV) & \rm unknown\\
& & & 0 & & &  $n+\bor9^{*}$ (2.788 MeV) & \rm unknown\\
& & & \rm unknown & & &  $n+\bor9^{*}$ (4.3 MeV) & \rm unknown\\
& & & 1 & & &  $n+\bor9^{*}$ (6.97 MeV) & \rm unknown\\
& & & 1 & & & $p + \be9$& \rm unknown\\
& & & 2 & & & $p+\be9^{*}$ (1.684 MeV)& \rm unknown\\
& & & 1 & & & $p+\be9^{*}$ (2.4294 MeV)& \rm unknown\\
& & & 1 & & & $p+\be9^{*}$ (2.78 MeV)& \rm unknown\\
& & & 0 & & & $p+\be9^{*}$ (3.049 MeV)& \rm unknown\\
& & & 0 & & & $p+\be9^{*}$ (4.704 MeV)& \rm unknown\\
& & & 1 & & & $p+\be9^{*}$ (5.59 MeV)& \rm unknown\\
& & & 1 & & & $p+\be9^{*}$ (6.38 MeV)& \rm unknown\\
& & & 2 & & & $p+\be9^{*}$ (6.76 MeV)& \rm unknown\\
& & & 1 & & & $p+\be9^{*}$ (7.94 MeV)& \rm unknown\\
& & & 1 & & & $p+\be9^{*}$ (11.283 MeV)& \rm unknown\\
& & & 1 & & & $p + \be9^{*}$ (11.81 \rm MeV)& \rm unknown\\
& & & 2 & & & $d + \be8$ &\rm unknown\\
& & & 0 & & & $d + \be8^{*}$ (3.03 MeV)&\rm unknown\\
& & & 2 & & & $d + \be8^{*}$ (11.35 MeV)&\rm unknown\\
& & & 1 & & & $\he3 + \li7$ &\rm unknown\\
& & & 1 & & & $\he3 + \li7^{*}$(0.47761 MeV) &\rm unknown\\
& & & 2 & & & $\alpha + \li6$ & \rm unknown\\
& & & 2 & & & $\alpha + \li6^{*}$ (2.186 MeV) & \rm unknown\\
& & & 2 & & & $\alpha + \li6^{*}$ (3.56 MeV) & \rm unknown\\
& & & 0 & & & $\alpha + \li6^{*}$ (4.31 MeV) & \rm unknown\\
& & & 0 & & & $\alpha + \li6^{*}$ (5.37 MeV) & \rm unknown\\
& & & 2 & & & $\alpha + \li6^{*}$ (5.65 MeV) & \rm unknown\\
& & & 1 & & & t (elastic) & \rm unknown\\
\hline
\hline
\end{tabular}
}
}
\end{center}
\caption{As in Table \ref{table:candidatesbebor8} listing the ground and excited final 
state channels for the 18.80 MeV energy level in \bor{10} 
for the $\be7+t$ initial state.   } 
\label{table:candidatesbor10d}
\end{table}%

\begin{table}[h!]
\begin{center}
\rotatebox{0}{
\scalebox{0.8}{
\begin{tabular}{|c|c|c|c|c|c|c|c|}
\hline
\hline
Compound Nucleus, & Initial State & $L_{\rm init}$ & $L_{\rm fin}$ & $E_{\rm res}$ & $\gtot$ & \rm Exit & \rm Exit Channel \\ 
$J^\pi, E_{\rm ex}$& & & & & &  Channels & Width \\
\hline
\bor{10}, $2^{-}$, 19.29 MeV & $\be{7} + t$ & 0 & \rm unknown & 620.9 $\rm{keV}$ & $190\pm20\ \rm{keV}$ &$\gamma $ &\rm unknown\\
& & & 0 & & & $n+\bor9$ & \rm unknown\\
& & & \rm unknown & & &  $n+\bor9^{*}$ (1.6 MeV) & \rm unknown\\
& & & 0 & & &  $n+\bor9^{*}$ (2.361 MeV) & \rm unknown\\
& & & 2 & & &  $n+\bor9^{*}$ (2.75 MeV) & \rm unknown\\
& & & 1 & & &  $n+\bor9^{*}$ (2.788 MeV) & \rm unknown\\
& & & \rm unknown & & &  $n+\bor9^{*}$ (4.3 MeV) & \rm unknown\\
& & & 0 & & & p + $\be9$& \rm unknown\\
& & & 1 & & & $p+\be9^{*}$ (1.684 MeV)& \rm unknown\\
& & & 0 & & & $p+\be9^{*}$ (2.4294 MeV)& \rm unknown\\
& & & 2 & & & $p+\be9^{*}$ (2.78 MeV)& \rm unknown\\
& & & 1 & & & $p+\be9^{*}$ (3.049 MeV)& \rm unknown\\
& & & 1 & & & $p+\be9^{*}$ (4.704 MeV)& \rm unknown\\
& & & 0 & & & $p+\be9^{*}$ (5.59 MeV)& \rm unknown\\
& & & 2 & & & $p+\be9^{*}$ (6.38 MeV)& \rm unknown\\
& & & 3 & & & $p+\be9^{*}$ (6.76 MeV)& \rm unknown\\
& & & 0 & & & $p+\be9^{*}$ (7.94 MeV)& \rm unknown\\
& & & 2 & & & $p+\be9^{*}$ (11.283 MeV)& \rm unknown\\
& & & 0 & & & $p + \be9^{*}$ (11.81 \rm MeV)& \rm unknown\\
& & & 1 & & & $d + \be8$ &\rm unknown\\
& & & 1 & & & $d + \be8^{*}$ (3.03 MeV)&\rm unknown\\
& & & 3 & & & $d + \be8^{*}$ (11.35 MeV)&\rm unknown\\
& & & 0 & & & $\he3$ &\rm unknown\\
& & & 1 & & & $\alpha + \li6$ & \rm unknown\\
& & & 1 & & & $\alpha + \li6^{*}$ (2.186 MeV) & \rm unknown\\
& & & 1 & & & $\alpha + \li6^{*}$ (4.31 MeV) & \rm unknown\\
& & & 1 & & & $\alpha + \li6^{*}$ (5.37 MeV) & \rm unknown\\
& & & 1 & & & $\alpha + \li6^{*}$ (5.65 MeV) & \rm unknown\\
& & & 0 & & & t (elastic) & \rm unknown\\
\hline
\hline
\end{tabular}
}
}
\end{center}
\caption{As in Table \ref{table:candidatesbebor8} listing the ground and excited final 
state channels for the 19.29 MeV energy level in \bor{10} 
for the $\be7+t$ initial state.} 
\label{table:candidatesbor10e}
\end{table}%

\begin{table}[h!]
\begin{center}
\rotatebox{0}{
\scalebox{0.75}{
\begin{tabular}{|c|c|c|c|c|c|c|c|}
\hline
\hline
Compound Nucleus, & Initial & $L_{\rm init}$ & $L_{\rm fin}$ & $E_{\rm res}$ & $\gtot$ & \rm Exit & \rm Exit Channel \\ 
$J^\pi, E_{\rm ex}$& State & & & & &  Channels & Width \\
\hline
\car{10},  & $\be{7} + \he3$ & \rm unknown & \rm unknown & \rm unknown & \rm unknown & p & \rm unknown \\
\rm unknown & & & \rm unknown & ($Q = 15.003$ MeV) & & $\alpha$ &\rm unknown\\
 & & &  &  & & \he3 (elastic) &  unknown\\
\hline
\bor{11}, & $\li{7} + \alpha$ & 0 & 1 & -103.7 $\rm{keV}$ & $ 1.346\ \rm{eV}$ &$\gamma$ (ground state)&$0.53\pm0.05$ \rm eV\\
$(3/2^{-})$, 8.56 MeV & & & 1 & & & $\gamma$ (2.125 MeV) &$0.28\pm0.03$\ \rm eV\\
& & & 1 & & & $\gamma$ (4.445 MeV) &$(4.7\pm1.1)\times10^{-2}$\ \rm eV \\
& & & 1 & & & $\gamma$ (5.020 MeV) &$(8.5\pm1.2)\times10^{-2}$\ \rm eV\\
& & & 1 & & & $\alpha$ (elastic) & \rm Unknown\\
\hline
\bor{11},& $\li{7} + \alpha$ & 2 & 1 & 256.3 $\rm{keV}$ & $ 4.37\pm0.02$\ \rm{eV} &$\gamma$ (ground state) &$4.10\pm0.20$ \rm eV \\
 $(5/2^{-})$, 8.92 MeV & & & 2 & & & $\gamma$ (ground state) &$(5.0\pm3.6)\times10^{-2}$\ \rm eV\\
 (Included)& & & 1 & & & $\gamma$ (4.445 MeV) &$0.22\pm0.02$\ \rm eV\\
& & & 1 & & & $\alpha$ (elastic) &\rm Unknown \\
\hline
\bor{11},& $\li{7} + \alpha$ & 3 & 1 & 526.3 $\rm{keV}$ & $ 1.9^{+1.5}_{-1.1}$\ \rm{eV} &$\gamma$ (ground state) &$(2.7\pm1.2)\times10^{-3}$\rm eV \\
 $7/2^{+}$, 9.19 MeV & & & 2 & & & $\gamma$ (4.445 MeV) &$0.25\pm0.09$\ \rm eV\\
& & & 0 & & & $\gamma$ (6.743 MeV) &$(3.8\pm1.3)\times10^{-2}$\ \rm eV\\
& & & 1 & & & $\alpha$ (elastic) &\rm Unknown \\
\hline
\bor{11},& $\li{7} + \alpha$ & 1 & 1 & 606.3\ \rm{keV} &  4\ \rm{keV} &$\gamma$ (ground state) &0.212 \rm eV \\
 $5/2^{+}$, 9.271 MeV & & & 0 & & & $\gamma$ (4.445 MeV) &0.802\ \rm eV\\
& & & 0 & & & $\gamma$ (6.743 MeV) & 0.137\ \rm eV\\
& & & 1 & & & $\gamma$ (6.792 MeV) & $<$ 0.007\ \rm eV\\
& & & 1 & & & $\alpha$ (elastic) &$\approx$\ 4 \rm{keV} \\
\hline
\car{11},& $\be{7} + \alpha$ & 1 & 1 & -43.3\ $\rm{keV}$ & $0.0105\ \rm{eV}$ &$\gamma$ (ground state) &\rm Unknown\\
 $3/2^{+}$, 7.4997 MeV & & & 0 & & & $\gamma$ (2.0 MeV) & \rm Unknown\\
& & & 1 & & & $\alpha$(elastic) & \rm Unknown\\
\hline
\car{11},& $\be{7} + \alpha$ & 0 & 1 & 557 \ $\rm{keV}$ & $11\pm7\ \rm{eV}$ &$\gamma$ (ground state) &$0.26\pm0.06$\ \rm eV\\
 $(3/2^{-})$, 8.10 MeV & & & 1 & & & $\gamma$ (2.0 MeV) & $(9.1\pm2.3)\times10^{-2}$\ \rm eV\\
& & & 0 & & & $\alpha$(elastic) & \rm Unknown\\
\hline
\hline
\end{tabular}
}
}
\end{center}
\caption{As in Table \ref{table:candidatesbebor8} listing resonances in \car{10}, \bor{11} and \car{11}. }
\label{table:candidatescar10bor11car11}
\end{table}%

The $\bor{10}$ compound nucleus has 
energy levels at 18.2, 18.43, 18.80 and 19.29 MeV, 
which we investigate. The 18.2 MeV level is uncertain 
experimentally \cite{tunlbor10} as indicated in Table~\ref{table:candidatesbor10a},  
and hence ideal for parametrizing. There is a 
\he3 entrance channel a little over 400 \rm keV below 
this level. The current total experimental width is 1.5 MeV 
which is very large and the branching ratios are unknown. 
The current uncertainty in the $\eres$ is 200 \rm keV. However, according 
to the plot in Fig.~\ref{fig:7A3Hecontours}, even a 200 \rm keV 
reduction in $\eres$ would not be sufficient to cause any appreciable 
destruction of \li7 as this reaction has negligible effect 
on the mass-7 abundance.
This is another illustration of the fact that reactions involving 
direct destruction of \li7 are unimportant.

The 18.43 MeV level
is better understood \cite{Yan:2002bc} and with a resonance 
energy of $\sim$ 640 \rm keV for the $\li7+\he3$ initial state
(Table \ref{table:candidatesbor10b})
and a total width of 340 \rm keV has a lower entrance probability and 
therefore is likely to be ineffective. This is 
evident from Fig.~\ref{fig:7A3Hecontours}. This level is also a 
sub-threshold resonance for the $\be7+t$ state 
(Table \ref{table:candidatesbor10c}),
with resonance energy, $\eres = -239.1 \ \rm keV$.  This is far below threshold
rendering it ineffective. 

\begin{figure}[htbp]
\begin{center}
\rotatebox{90}{
\scalebox{0.8}{
\includegraphics[width=1.0\textwidth]{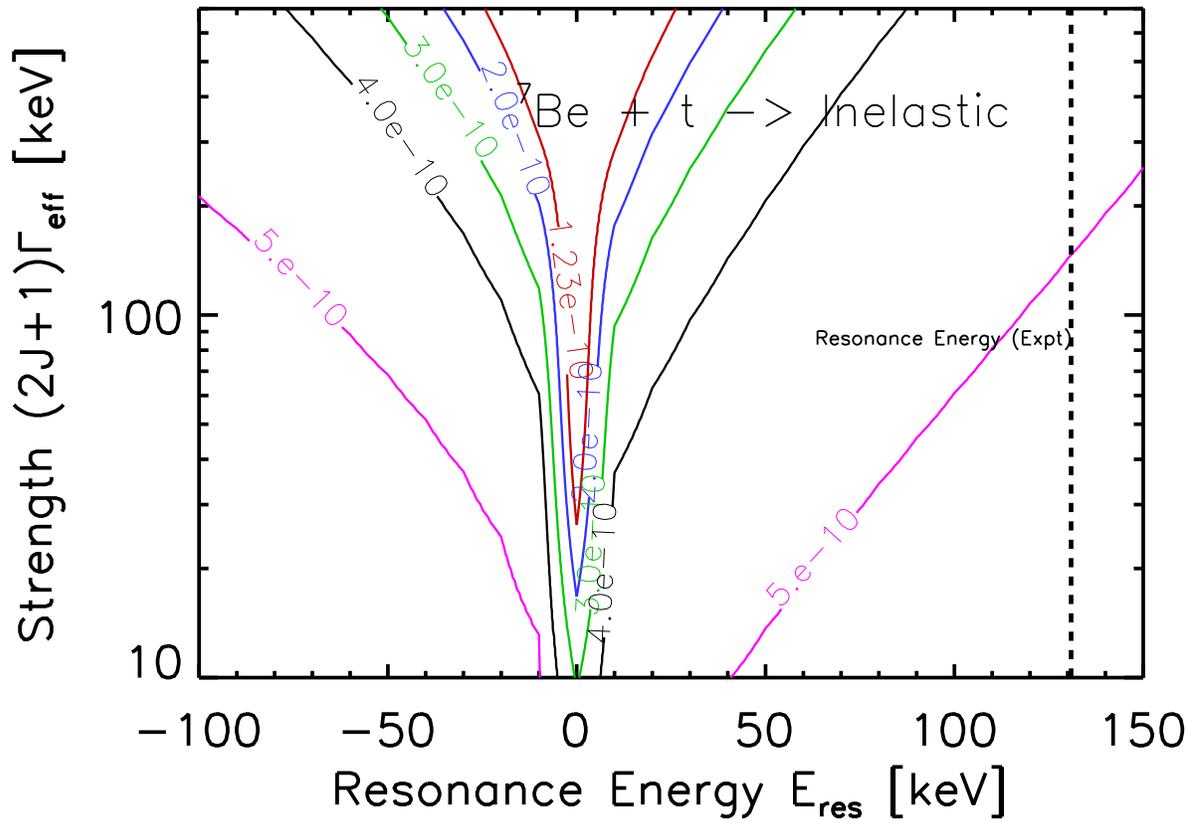}
}
}
\caption{As in Fig. \ref{fig:8Becontours} for the resonances in the \bor{10} compound nucleus involving initial states \be{7}$+t$. }
\label{fig:10Bcontours}
\end{center}
\end{figure}

\begin{figure}[htbp]
\begin{center}
\rotatebox{90}{
\scalebox{0.8}{
\includegraphics[width=\textwidth]{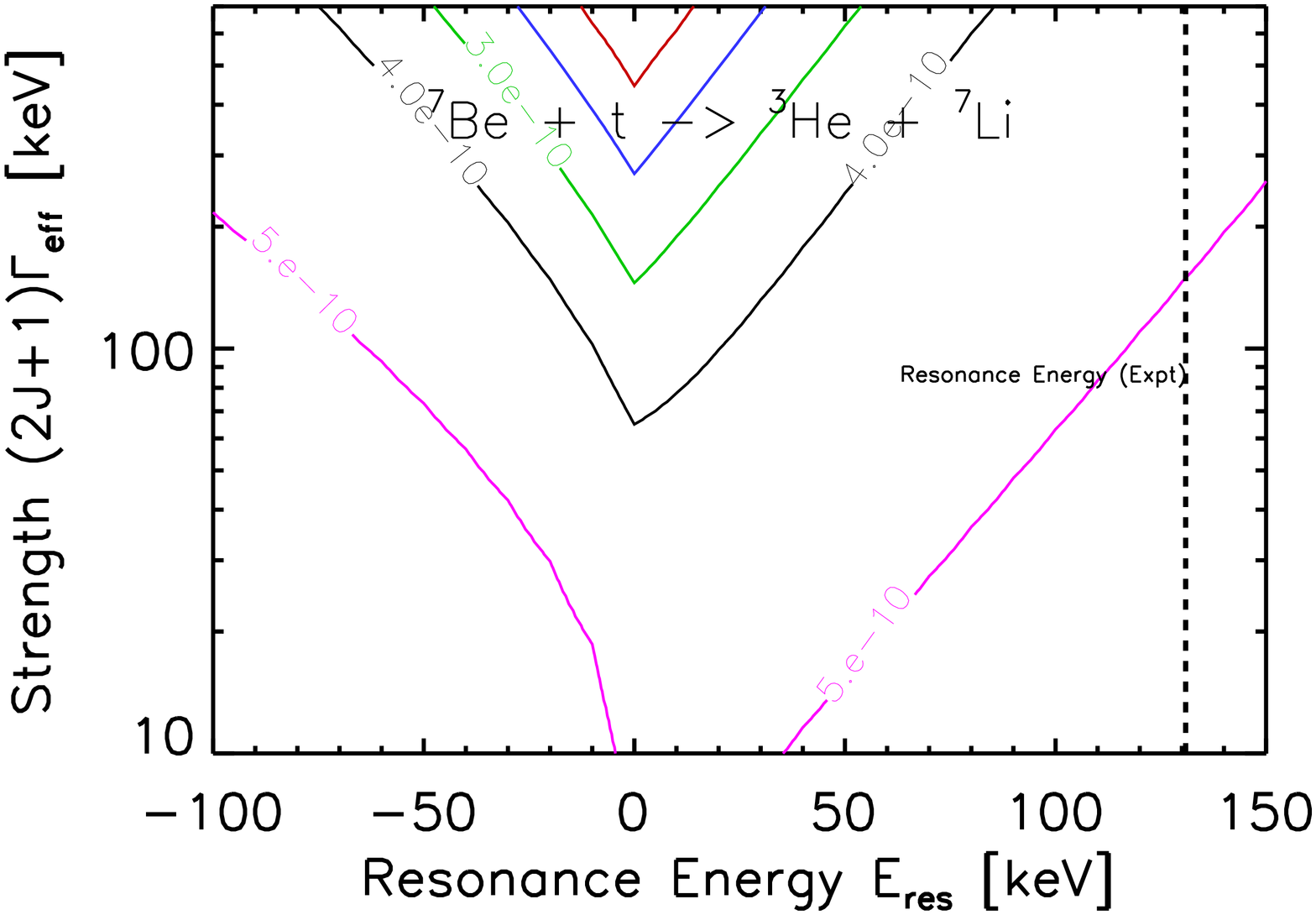}
}
}
\caption{As in Fig. \ref{fig:8Becontours} for the resonances in the reaction \be{7}(t,\he{3})\li{7}. }
\label{fig:7bet3hecontours}
\end{center}
\end{figure}

Staying with $\be{7}+t$, the closest 
energy level above the entrance energy of 18.669 MeV is the 
18.80 MeV $(2^{+})$ level (Table \ref{table:candidatesbor10d}), 
which corresponds to a resonance 
energy of $\approx$ 130 \rm{keV} \cite{tunlbor10}. The exit channel 
widths for $p$ and \he{3} are unknown experimentally and 
thus, this is a candidate for parametrization. 
There is a weak upper limit on $\gtot < 600 \ \rm keV$ \cite{tunlbor10},
which for $J_*=2$ is off scale in Figs.~\ref{fig:10Bcontours}
and \ref{fig:7bet3hecontours}.
The contour plot in Fig.~\ref{fig:10Bcontours} show 
that for a central value of resonance energy 
of $\approx$ 130 \rm{keV} shown by the vertical dashed line, 
resonance strength of 
just under 1 MeV is required which is very high. 
Also, 
parity requirements force $L =1$, which will cause suppression 
of this channel. We note that there is no quoted uncertainty for this energy
level and neighboring levels have typical uncertainties of 100-200 keV.
Therefore it may be possible (within 1-2$\sigma$) that the state lies at 
an energy of 100 keV lower and would energetically, 
have a chance at solving the \li7 problem.
This is true for the $p$ exit channel.

The \he{3} exit channel may also reduce mass 7,  through the formation 
of the \li{7} which is much easier to destroy. This is reflected 
in Fig.~\ref{fig:7bet3hecontours}, which shows that at resonance 
energies of $\leq 100$ keV, a 
strength of a few 100 keV, but less than 600 keV 
may be sufficient to achieve comparable 
destruction of \be{7} as the 16.71 MeV resonance. 
The caveat is that for such values of strengths, the narrow 
resonance approximation does not hold true and this may lead 
to a reduced effect. Nevertheless, this is yet another 
case deserving a detailed comparison with the theoretical 
limits which will follow in \S~\ref{reducedlist}. 
Once again, definitive conclusions can be 
drawn only based on experimental data.

The 19.29 MeV level (Table \ref{table:candidatesbor10e})
is energetically harder to access and 
with a total width of only 190 \rm keV, it is unlikely to 
be of significance, despite being less studied.

\begin{figure}[h!]
\begin{center}
\rotatebox{90}{
\scalebox{0.8}{
\includegraphics[width=\textwidth]{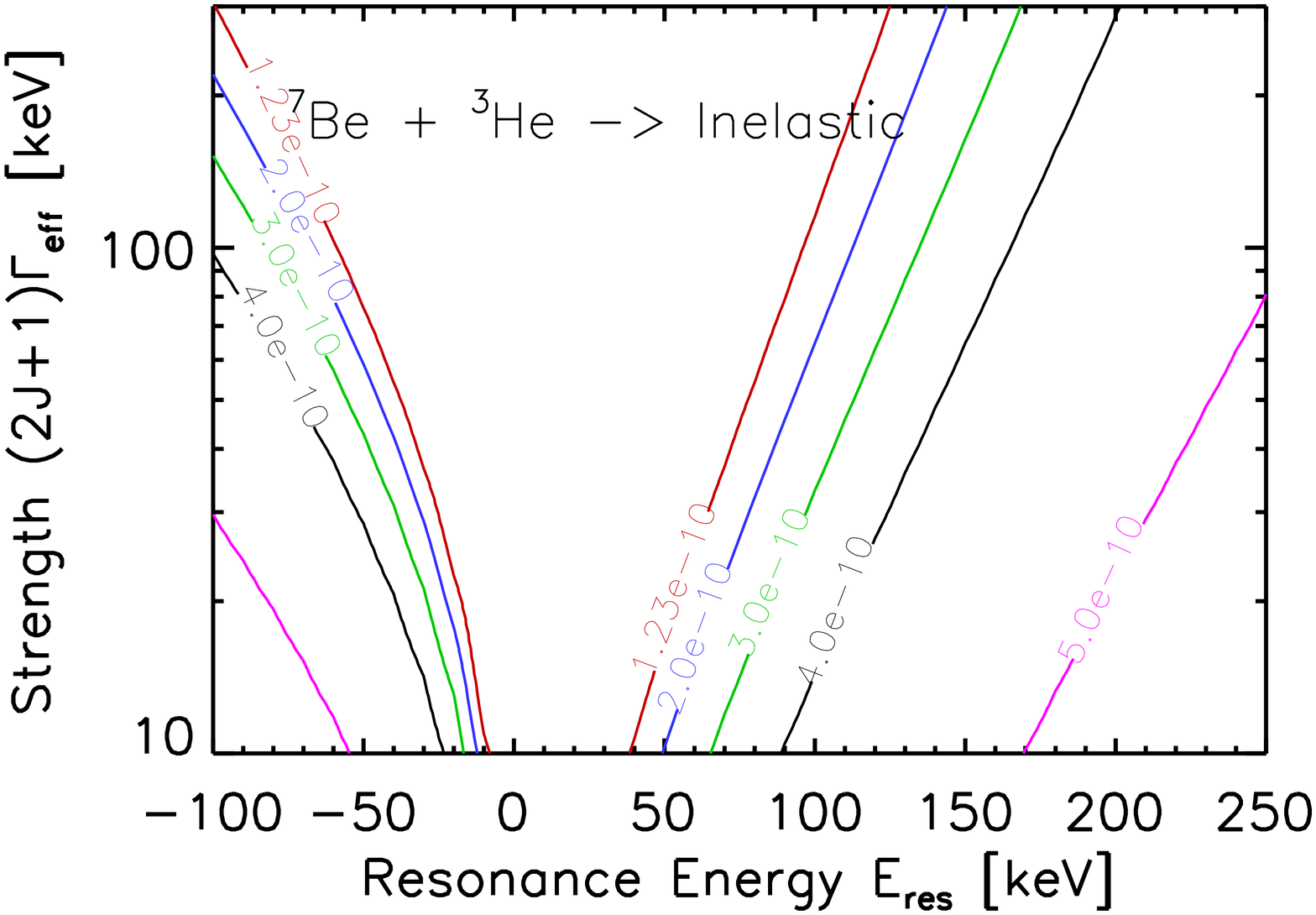}
}
}
\caption{As in Fig. \ref{fig:8Becontours} for the resonances in \car{10} involving initial state \be{7} + \he{3}}
\label{fig:7be3hecontours}
\end{center}
\end{figure}

The \car{10} nucleus \cite{tunlcar10} 
appearing in Table \ref{table:candidatescar10bor11car11} shows
large uncertainties and experimental gaps at higher energy levels 
which may be relevant to entrance channels involving \be7. 
Reactions involving the $\be7+\he3$ initial state 
could contribute in destroying \be7 if 
there exists a resonance in the parameter space  
shown in the 
Fig.~\ref{fig:7be3hecontours}. 
These reactions win over 
those involving the $\be7 + t$ state, because 
\he3 is substantially more abundant than $t$, 
but are worse off due to a higher Coulomb barrier.
The entrance energy for \be7 + \he3 is 15.0 MeV.
As one can see from the figure,
a 1$^-$ or 2$^-$ state with a resonance energy of
either -10 keV or 40 keV
corresponding to 
energy levels of 14.99 and 15.04 MeV respectively 
with a strength as high as a few 10's of keVs is what it will take 
to solve the lithium problem with this initial state.
Thus, any \car{10} resonance near these energies which 
may have been missed by experiment
may be interesting as a solution to the lithium problem;
we return to this issue in more detail in \S~\ref{reducedlist}.

\subsection{\label{compnuc11} A = 11 Compound Nucleus}

For $\bor{11}$ \cite{tunlbor11}, Table \ref{table:candidatescar10bor11car11}
shows that the entrance channel, 
$\li{7}+\alpha$ is at 8.6637 MeV which is 103.7 $\rm{keV}$ 
above the resonant energy level at 8.560 MeV and 
$\approx 260\ \rm{keV}$ below the resonant energy level 
at 8.92 MeV.  Parity demands 
angular momentum to be 0. 
Both states are at relatively large $|E_{\rm res}|$
and are not capable of making a sizable impact on the 
\li7 abundance. 
Table \ref{table:candidatescar10bor11car11} further lists states 9.19 MeV (which 
requires $L = 3$ and has a total width of $< 2$ eV) and 9.271 MeV (whose 
decay is dominated by the elastic channel) which have progressively larger resonant energies
and are unlikely to provide a solution.

For $\car{11}$ \cite{tunlcar11}, 
the entrance channel, $\be{7}+\alpha$ is at 7.543 MeV 
which is 43 $\rm{keV}$ above the resonant energy level 
at 8.560 MeV and $557\ \rm{keV}$ below the resonant 
energy level at 8.10 MeV. 

\begin{figure}[htbp]
\begin{center}
\rotatebox{90}{
\scalebox{0.8}{
\includegraphics[width=1.0\textwidth]{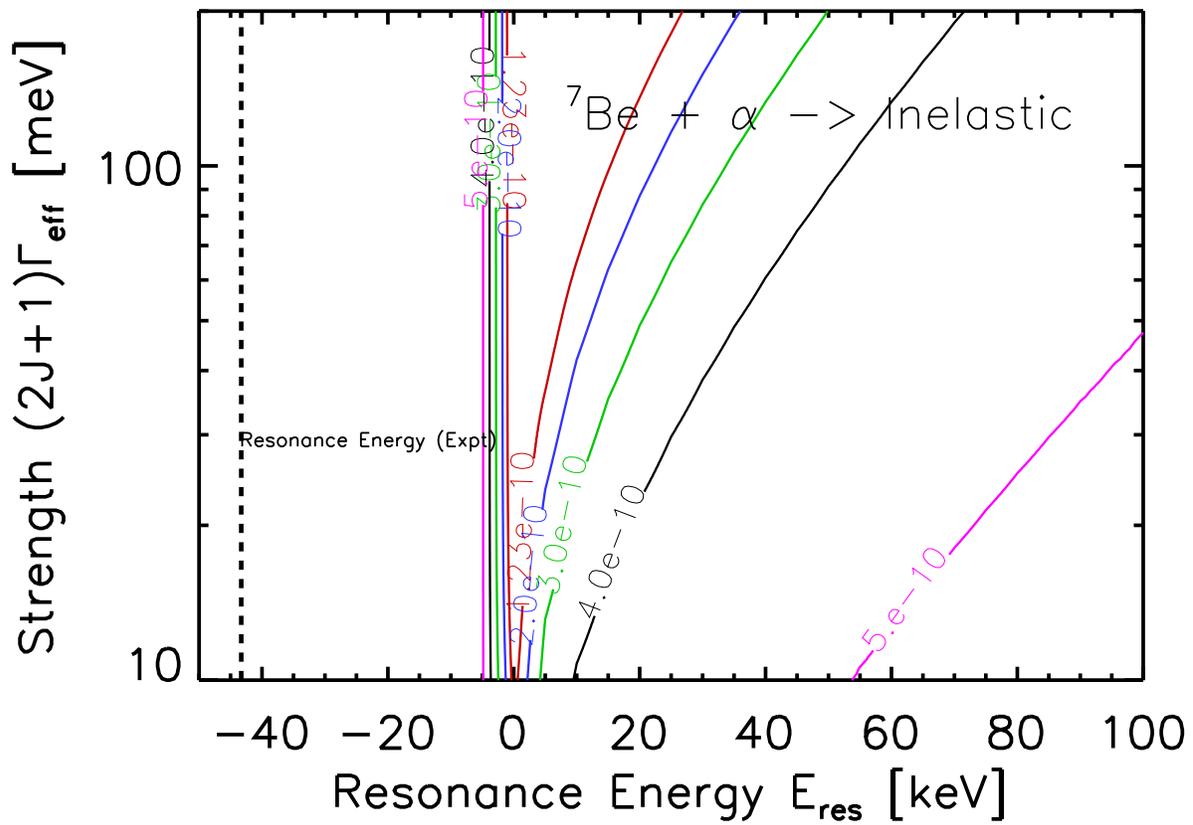}
}
}
\caption{As in Fig. \ref{fig:8Becontours} for the resonances in \car{11} involving initial states \be{7}$+\alpha$.}
\label{fig:11Ccontours}
\end{center}
\end{figure}

As seen in Fig.~\ref{fig:11Ccontours}, we find that the 
sub-threshold resonance in the $\car{11}$ nucleus, 
produces a very insignificant effect on $\be{7}$ in agreement with the 
claim in \cite{Cyburt:2009cf}. 
The super-threshold resonance states 
are also too far away at resonance energies, 
$557\ \rm{keV}$ and $260\ \rm{keV}$ for 
$\be{7}(\alpha,\gamma)\car{11}$ and 
$\li{7}(\alpha,\gamma)\bor{11}$ respectively.

However, Fig.~\ref{fig:11Ccontours} shows that the presence of a 
(missed) resonance at resonance energies of few tens of keV 's, 
requires a very meagre strength of the order of tens of meV 's to 
destroy mass 7 substantially. Strengths of this order are typical of 
electromagnetic channels. It is difficult to assess the probability
that a $^{11}$C state at 7.55 MeV has been overlooked.

\section{Reduced List of Candidate Resonances \label{reducedlist} }

Having systematically identified all possible known 
resonant energy levels which could affect BBN,
we find most of these levels are ruled out immediately as
promising solutions, based on their measured locations,
strengths, and widths. 
As expected, the existing 
electromagnetic channels are too weak to cause significant 
depletion of lithium owing to their small widths. 

From amongst the various hadronic channels listed 
in the tables above, we have seen that all channels are 
unimportant except three which are summarized in 
table~\ref{table:remcandidates}. 
The $\be7 + d$ channels involving the 
16.71 MeV resonance in \bor9, the  $\be7 + t$ channels involving the 
18.80 MeV resonance in \bor{10} and $\be7 + \he3$ channels.
These are ones where a more detailed theoretical calculation of 
widths is required to decide whether they may be important 
or not. For each reaction,
the Wigner limit, eq.~(\ref{eq:TeichmannWignerlimit}), to 
the reduced width $\gamma^{2}$ imposes a bound on 
$\Gamma_{L}$ via eq.~(\ref{eq:partialwidth}).
Specifically, the penetration factor, $P_{L}(E,a)$, 
must be estimated to see if the required strengths according 
to Figs.~\ref{fig:7bedpcontours}, ~\ref{fig:10Bcontours} 
~\ref{fig:7bet3hecontours}, and ~\ref{fig:7be3hecontours} 
to solve the problem are at all attainable. The penetration factor is given by
\beq
\label{eq:penetrationfactor}
P_{L}(E,a) = \frac{1}{G^{2}_{L}(E,a) + F^{2}_{L}(E,a)}
\eeq 
where $G_{L}(E,a)$ and $F_{L}(E,a)$ are 
Coulomb wavefunctions.

We note that the Coulomb 
barrier penetration factor decreases as 
the energy of the projectile and/or the channel radius, $a$, increases. 
For a narrow resonance, 
the relevant projectile energy is $E \approx \eres$, 
which is set by nuclear experiments (where available) 
and their uncertainties. The channel radius corresponds 
to the boundary between the compound nucleus in 
the resonant state and the outgoing / incoming particles. 
Therefore, the channel radius depends on the properties 
of the compound state and the particles into which it decays. 

Consider the case of $\be7 + d$, which has resonance energy, 
$\eres = 220\pm100\ {\rm keV}$ and initial angular 
momentum, $L_{init} = 1$. A naive choice for the 
channel radius is the ``hard-sphere'' approximation,
\beqa
\label{eq:hardsphere}
a_{12} = 1.45\ (A_{1}^{1/3} + A_{2}^{1/3}) {\rm fm}
\eeqa
which gives $a_{27} = 4.6\ \rm fm$.
Using the Coulomb functions, 
$\Gamma_{1}$ is of order a few keVs.
The corresponding strength, $\geff$ should be essentially the same 
and we further gain a factor of 6 from the spin of this state. 
This suggests by using figure~\ref{fig:7bedpcontours}, 
that this resonance should fall short of the
width required to solve or even 
ameliorate the problem. 

However,  reactions 
involving light nuclides including $A = 7$ are found 
to have channel radii exceed the hard-sphere approximation \cite{Cyburt:2009cf}. 
We thus consider larger radii and find that for values  
higher than around 10 fm, we get a width 
which has the potential to 
change the \li7 abundance noticeably.
The Wigner limit
\beqa
\label{eq:wignerscaling}
a^{2} = \frac{3 \hbar^{2}}{2\ \mu\ \eres}
\eeqa
gives a larger radius, $a_{27} = 13.5$ fm, 
which gives one a better chance of solving the problem. This 
is consistent with the conclusions drawn by \cite{Cyburt:2009cf}.

For the $\be7 + t$ initial state, the 18.80 MeV state of \bor{10} has a resonance energy
$E = 0.131\ \rm MeV$ and $L_{init} = 1$. There is no 
experimental error bar on the resonance energy. 
The hard sphere approximation gives 
$a_{37} = 4.9\ \rm fm$. 
This gives a 
width, $\Gamma_{1}$, which is less than a tenth of a keV,
and is orders of magnitude lower than what is needed. 
In the spirit of what we did in the earlier case, 
using eq.~(\ref{eq:wignerscaling}) gives 
a channel radius, $a_{37} = 15\ {\rm fm}$ 
improving the situation 
by almost 2 orders of magnitude in $\Gamma_1$. 
If, in addition to increasing 
$a_{37}$,  the resonance energy were to be 
higher by 100 $\rm keV$, 
then $\Gamma_{1}$ could be large enough 
to change the \li7 abundance noticeably.

The $\be7 + \he3$ initial state will have $\car{10}$
as the compound state. The structure of the \car{10} nucleus is 
not well studied
experimentally \cite{tunlcar10} nor theoretically.
In particular, we are unaware of any published data
on \car{10} states near the $\be7 + \he3$
entrance energy, i.e., states at or near 
$E_{\rm ex}(\car{10}) \approx Q(\be7 + \he3) = 15.003$ MeV. 
To our knowledge, there has not been any search for
narrow states in this region.
The potential exit channels of importance are 
$\bor{9}+p$ and $\be6+\alpha$. Because $J^\pi(\he3)=1/2^+$
$J^\pi(\be7)=3/2^-$, to have $L_{\rm init}=0$ and
thus no entrance angular momentum barrier would require
the \car{10} state to have \
\beq
J^\pi = (\mbox{1  or 2})^- \ \ 
\eeq
Because $J^\pi(\bor{9}) = 3/2^{-}$, the entrance
channel spin and parity required to give $L_{\rm init} = 0$
will also allow $L_{\rm fin} = 0$ for the $\bor{9}+p$.
On the other hand,
in the final state $\be6+\alpha$
both \be6 and \he4 have $J^\pi = 0^+$.
Thus if the putative \car{10} state has $J^\pi = 1^-$,
this forces the $\be6+\alpha$ final state to have $L_{\rm fin} = 1$,
and thus this channel will be suppressed by an angular momentum
barrier relative to $\bor{9}+p$.

Using 
eq.~(\ref{eq:hardsphere}),
we again get $a_{37} = 4.9 \ \rm fm$. 
Taking $L_{init} = 0$ and $E = 0.2\ \rm MeV$,  
$\Gamma_{0}$ is about $10^{-3}$ \rm keV and
is extremely small. 
However, the penetration factor is highly 
sensitive to the channel radius and a relatively small increase in $a$ 
increases the width by orders of magnitude. 
Increasing the energy does reduce the 
penetration barrier, but a higher width is required 
due to thermal suppression. In order to get a sizable width, which is required to 
solve the problem according to figure~\ref{fig:7be3hecontours}, 
$a_{37}$must be $\ga 30\ $ \rm fm. At this energy, 
this radius is somewhat larger than 
what is afforded by Eq.~\ref{eq:wignerscaling}. 


\begin{table}[h!]
\begin{center}
\rotatebox{0}{
\scalebox{0.70}{
\begin{tabular}{|c|c|c|c|c|c|c|c|}
\hline
\hline
Compound Nucleus, & Initial & $L_{\rm init}$ & $L_{\rm fin}$ & $E_{\rm res}$ & $\gtot$ & \rm Exit & \rm Exit Channel \\ 
$J^\pi, E_{\rm ex}$& State & & & & &  Channels & Width \\
\hline
\bor{9}, $(5/2^{+})$, 16.71\ \rm MeV & $\be{7} + d$ & 1 & 0 & 219.9 $\rm{keV}$ & \rm unknown & $p + \be{8}^{*}$ (16.63 \rm MeV) &\rm unknown\\
& & & 1 & & & $\alpha + \li5$ &\rm unknown\\
\hline
\bor{10},& $\be{7} + t$ & 1 & 1 & 130.9 $\rm{keV}$ & $< 600\ \rm{keV}$ &$p + \be9^{*}$ (11.81 MeV) &\rm unknown\\
 $2^{+}$, 18.80 MeV & & & 1 & & & $\he3$ &\rm unknown\\
& & & 2 & & & $\alpha$ &\rm unknown\\
\hline
\car{10},  & $\be{7} + \he3$ & \rm unknown & \rm unknown & \rm unknown & \rm unknown & p & \rm unknown \\
\rm unknown & & & \rm unknown & ($Q = 15.003$ MeV) & & $\alpha$ &\rm unknown\\
 & & &  &  & & \he3 (elastic) &  unknown\\
\hline
\hline
\end{tabular}
}
}
\end{center}
\caption{This table lists surviving candidate resonances.} 
\label{table:remcandidates}
\end{table}%

\section{Discussion and Conclusions}
\label{disc}

The lithium problem was foreshadowed before precision 
cosmic microwave background data,
was cast in stark light by the first-year WMAP results,
and has only worsened since.
While astrophysical solutions are not ruled out, they
are increasingly constrained.  Thus, a serious
and thorough evaluation of all possible nuclear physics
aspects of primordial lithium production is urgent
in order to determine whether the lithium problem
truly points to new fundamental physics.

Reactions involving the primordial {\em production} of mass-7,
and its lower-mass progenitor nuclides, are very well studied
experimentally and theoretically, 
and leave no room for surprises at the level needed to solve
the lithium problem \cite{Cyburt:2009cf,boyd,bbn2}.
Lithium {\em destruction} reactions are
less well-determined. While the dominant destruction channels
$\be7(n,p)\li7$ and $\li7(p,\alpha)\alpha$ have been
extensively studied, in contrast, the subdominant destruction channels
are less well-constrained.

We therefore have exhaustively cataloged possible
resonant, mass-7 destruction channels.
As evidenced by the large size of 
Tables \ref{table:candidatesbebor8}--\ref{table:candidatescar10bor11car11},
the number of potentially interesting compound states is
quite large.
However, it is evident that the basic conservation laws such as angular 
momentum and parity coupled with the requirement of resonant 
reactions to be 2--3 times the $\be7(n,p)\li7$ rate 
prove to be extremely restrictive on the options for a resonant 
solution to the lithium problem, and reduces the possibilities
dramatically.

Given existing nuclear data, 
there are several choices for experimentally identified 
nuclear resonances which come close to removing the 
discrepancy between the lithium WMAP+BBN predictions and observations as 
tabulated in \S~\ref{reducedlist}. The  
16.71 MeV level in \bor{9} compound nucleus, and the 18.80 MeV\ level
in the \bor{10} compound nucleus 
are two such candidates.
It is possible, however, that resonant effects have been
neglected in reactions passing through states which
have been entirely missed.  In all of the plots above, we have illustrated
the needed positions and strengths of such states,
if they exist. One possibility involving the compound state 
\car{10} is poorly studied experimentally, especially at 
higher energy states close to the Q-value for $\be7+\he3$.

Any of these resonances (or a combination) could offer a partial or complete
solution to the lithium problem, but in each case, we find that
large channel radii ($a > 10$ fm) are needed in order that
the reaction widths are large enough.
We confirm the results of Cyburt and Pospelov \cite{Cyburt:2009cf} 
in this regard concerning $\be7+d$, and we also find similar channel radii
are needed for $\be7+t$, while larger radii are required for $\be7+\he3$.
Obviously, nature need not be so kind (or mischievous!) in providing such
fortuitious fine-tuning.
But given the alternative of new physics solutions to the lithium problem,
it is important that all conventional approaches be exhausted.

Thus, based on our analysis, quantum mechanics 
could allow resonant properties that can remove or 
substantially reduce the lithium discrepancy. 
An experimental effort to measure the
properties of these resonances, however can conclusively 
rule out these resonances as solutions. If all possible resonances
are measured and found to be unimportant for BBN,
this together with other recent work \cite{boyd}, will remove 
any chance of a ``nuclear solution'' to the lithium problem, and
substantially increase the possibility of a new physics solution.
Thus, regardless of the outcome, experimental probes of
the states we have highlighted will complete the firm empirical foundation
of the nuclear physics of BBN and will make a crucial contribution
to our understanding of the early universe.

\bigskip

We are pleased to acknowledge useful and stimulating conversations with
Robert Wiringa, Livius Trache, Shalom Shlomo, Maxim Pospelov, Richard Cyburt,
and Robert Charity.
The work of KAO was supported in part by DOE grant
DE--FG02--94ER--40823 at the University of Minnesota.

\appendix

\section{The Narrow Resonance Approximation}
\label{sect:narrow}

Consider a reaction $A + b \rightarrow C^{*} \rightarrow c + D$,
which passes through an excited state of the compound nucleus $C^*$.
We treat separately normal and subthreshold reactions,
defined respectively by a positive and negative sign of
the resonance energy
$\eres = E_{\rm ex} - Q_C$,
where $E_{\rm ex}$ is the excitation energy 
of the $C^*$ state considered,
and $Q_C = \Delta(A)+\Delta(B)-\Delta(C^*)$.

In general, the thermally averaged rate is
\beq
\avg{\sigma v} = 
   \frac{\int d^3v \ e^{-\mu v^2/2T} \sigma v}
        {\int d^3v \ e^{-\mu v^2/2T}}
  = \sqrt{\frac{8}{\pi \mu}}  \
  T^{-3/2} \int_0^\infty dE \ E \ \sigma(E) e^{-E/T}
\eeq
For a Breit-Wigner resonance with widths not strongly varying with
energy, this becomes
\beq
\label{eq:narrowint}
\avg{\sigma v}
  = \frac{4\pi\omega \g{init} \g{fin}}{(2 \pi \mu T)^{3/2}} 
\int_0^\infty dE \frac{e^{-E/T}}
        { (E-\eres)^2 + (\gtot/2)^2 }
\eeq
Thus the thermal rate is controlled by the integral of the
Lorentzian resonance profile modulated with the exponential
Boltzmann factor.

The narrow resonance approximation has usually only
been applied to the normal resonance case,
and assumes that the
total resonance width is small compared to the temperature:
$\gtot \ll T$.

\subsection{Narrow Normal Resonances}

In the normal or ``superthreshold'' case, the integral
includes the peak of the Lorentzian where $E = \eres$.
The narrow condition then guarantees that over the Lorentzian width,
the Boltzmann factor does not change appreciably,
and so we make the approximation 
\beq
\exp\left(-\frac{E}{T}\right) \approx \exp\left(-\frac{\hat{E}}{T}\right)
\eeq
where we choose the ``typical'' energy to be the
peak of the Lorentzian, $\hat{E} = \eres$.
Then the integral becomes
\beq
\avg{\sigma v}_{\gtot \ll T}
  \approx \frac{\omega \g{init} \g{fin}}
               {2 (2 \pi \mu T)^{3/2}} 
  e^{-\eres/T}
\int_0^\infty dE \frac{1}
        { (E-\eres)^2 + (\gtot/2)^2 }
\eeq
Furthermore, it is usually also implicitly assumed
that the resonance energy 
is large compared to the width:  $\eres \gg \gtot$.
Then the integral gives $2\pi/\gtot$, and
the thermally averaged cross-section under this approximation
is given by \cite{Angulo:1999zz},
\begin{eqnarray}
\label{eq:narrow0}
\avg{\sigma v}_{\gtot \ll T, E_{\rm res}}  
  &=&
  \omega \ \geff \pfrac{2\pi}{\mu T}^{3/2} e^{-\eres/T}  \\
  & = &
  2.65\times10^{-13} \mu^{-3/2} \ \omega \ \geff \ T^{-3/2}_{9} 
   \exp(-11.605\ \eres / T_{9}) \ \rm{cm^{3} s^{-1}} 
\end{eqnarray}
where the latter expression has $T_9 = T/10^9$ K.

Note however, that eq.~(\ref{eq:narrowint})
is exactly integrable as it stands and does not
require we make the usual $\eres \gg \gtot$ approximation.
Thus for the normal case we modify the usual reaction rate
and instead adopt the form
\beq
\label{eq:normal}
\avg{\sigma v}_{\rm narrow,normal} =
\avg{\sigma v}_{\gtot \ll T, E_{\rm res}} \ f(2\eres/\gtot)
\eeq
Here we introduce a temperature-independent
correction for finite $\eres/\gtot$ (still with $\eres > 0$)
\beq
f(u) = \frac{1}{2} + \frac{1}{\pi}\arctan u \ \ .
\eeq
This factor spans $f \rightarrow 1/2$ for $\eres \ll \gtot$ to
$f \rightarrow 1$ for $\eres \gg \gtot$.

In practice, we adopt a slightly modified version of the correction
factor in our plots.
Recall that in Figs.~\ref{fig:8Bcontours}--\ref{fig:11Ccontours}, we show
results for lithium abundances in the presence of resonant
reactions with fixed input channels, but without
reference to a specific final state.
Without the correction factor,
the resonant reaction rate is characterized by two parameters,
$\eres$ and $\geff$.  These two parameters are insufficient to
specify the correction factor, which depends on $\eres/\gtot$.
Rather than separately introduce $\gtot$, we instead
approximate the correction factor as $f(2\eres/\geff)$.
Because $\geff < \gtot$ and $f$ is monotonically increasing, 
this always {\em underestimates} the value of $f$ and thus
conservatively {\em understates} the importance of the resonance
we seek (but the approximation is never off by more than a factor of 2
in the normal case).

\subsection{Narrow Subhreshold Resonances}

Still making the narrow resonance approximation $\gtot \ll T$, we
now turn to the subthreshold case, in which $\eres < 0$.
To make effect of the sign change explicit, we rewrite
eq.~(\ref{eq:narrowint}) as
\beq
\avg{\sigma v}
  = \frac{\omega \g{init} \g{fin}}{2(2 \pi \mu T)^{3/2}} 
\int_0^\infty dE \frac{e^{-E/T}}
        { (E+|\eres|)^2 + (\gtot/2)^2 }
\eeq
Now the integrand always excludes the resonant peak,
and only includes the high-energy wing.
As with the normal case, the narrowness of the resonance
implies that the Boltzmann exponential does not change much
where the Lorentzian has a significant contribution, and so
we again will approximate $e^{-E/T} \approx e^{-\hat{E}/T}$.
Since we avoid the resonant peak, the choice of $\hat{E}$
not as straightforward in the subthreshold case where
we took $\hat{E} = \eres$.
This choice makes no sense in the subthreshold
case, because the $e^{-\eres/T} > 1$ in the subthreshold case,
yet obviously kinetic energy $E>0$ and thus
the Boltzmann factor must always be a suppression and not an enhancement!

Yet clearly $|\eres|$ remains an important scale.
Thus we put $\hat{E} = \hat{u} |\eres|$, and we have examined
results for different values of the dimensionless parameter
$\hat{u}$.  We find good agreement with numerical results
when we adopt $\hat{u} \approx 1$, i.e., $\hat{E} = |\eres|$.
Thus for the subthreshold case
we adopt a reaction rate which is
in closely analogous to the normal case:
\beq
\label{eq:subth}
\avg{\sigma v}_{\rm narrow, \, subthreshold} =
\omega \ \geff \pfrac{2\pi}{\mu T}^{3/2} e^{-|\eres|/T}
  f\left(-2|\eres|/\gtot\right)
\eeq
Similarly to the normal case, as the reaction becomes
increasingly off-resonance, i.e., as $|\eres|$ grows,
there is an exponential suppression.
In addition, the 
correction factor has limits $f \rightarrow 1/2$ for $|\eres| \ll \gtot$,
and $f \rightarrow 0$ as $|\eres| \gg \gtot$.
Finally, note that, as a function of $\eres$,
our subthreshold and normal rates match at $\eres = 0$, as
they must physically.

\newpage

{}

\end{document}